\documentclass[aps,prc,twocolumn,superscriptaddress,nofootinbib,tightenlines,
preprintnumbers,showkeys,floatfix]{revtex4-2}

\usepackage[utf8]{inputenc}
\usepackage[english]{babel}
\usepackage{amssymb,amsthm,amsmath,amstext,amsbsy,amsopn}
\usepackage{bbm}
\usepackage{nicefrac}
\usepackage{slashed}
\usepackage{physics}
\usepackage{siunitx}
\usepackage{graphicx}
\usepackage{hyperref}
\usepackage[capitalise]{cleveref}
\usepackage{leftidx}
\usepackage{environ}
\usepackage{mathtools}
\usepackage{xspace}
\usepackage{array}
\usepackage{xcolor}
\usepackage{pgfplotstable}
\usepackage{float}
\usepackage{placeins}
\usepackage{isotope}
\usepackage{tikz}

\usetikzlibrary{shapes,patterns,decorations.markings}

\newcommand{\ie}{{i.e.}\xspace}

\newcommand{\etal}{\textit{et al.}\xspace}

\newcommand{\via}{{via}\xspace}

\newcommand{\MeV}{\ensuremath{\mathrm{MeV}}}
\newcommand{\fm}{\ensuremath{\mathrm{fm}}}

\DeclareMathOperator*{\argmin}{arg\,min}

\newcommand{\mbraket}[3]{\langle #1|#2|#3\rangle}

\newcommand{\ii}{\mathrm{i}}

\NewEnviron{subalign}[1][]{%
\begin{subequations}\begin{align}
  \BODY
\end{align}\label{#1}\end{subequations}
}

\NewEnviron{spliteq}{%
\begin{equation}\begin{split}
  \BODY
\end{split}\end{equation}
}

\newcommand{\cm}{\text{cm}}

\newcommand{\vecx}{\mathbf{x}}

\newcommand{\vecR}{\mathbf{R}}

\newcommand{\cdd}{{\cdot\cdot}}

\newcommand{\dvrsum}[2]{\sum\limits_{#1={-}#2/2}^{#2/2-1}}

\DeclareRobustCommand\shadingslash{
 \tikz
 \draw[blue!60,opacity=0.4,pattern=north east lines, pattern color=blue]
 (0.0,-0.0) rectangle (0.6,0.25);
}
\DeclareRobustCommand\shadingbackslash{
 \tikz%
 \draw[orange,opacity=0.4,pattern=north west lines, pattern color=orange]
 (0.0,-0.0) rectangle (0.6,0.25);
}
\DeclareRobustCommand\tikzdiamond{
 \tikz%
 \node[diamond,draw=black,fill=violet,line width=0.3mm,inner sep=0pt,
  minimum size=2mm
 ]
 () at (0.0,0.0) {};
}

\renewcommand{\vec}[1]{\mathbf{#1}}

\definecolor{ccolor}{named}{red}

\begin{document}

\title{Three-body resonances in pionless effective field theory}

\author{S.~Dietz}
\email{sebastian.dietz@physik.tu-darmstadt.de}
\affiliation{Technische Universität Darmstadt, Department of Physics,
64289 Darmstadt, Germany}

\author{H.-W.~Hammer}
\email{hans-werner.hammer@physik.tu-darmstadt.de}
\affiliation{Technische Universität Darmstadt, Department of Physics,
64289 Darmstadt, Germany}
\affiliation{ExtreMe Matter Institute EMMI and Helmholtz Forschungsakademie
  Hessen f\"ur FAIR (HFHF),
GSI Helmholtzzentrum für Schwerionenforschung GmbH,
64291 Darmstadt, Germany}

\author{S.~König}
\email{skoenig@ncsu.edu}
\affiliation{Department of Physics, North Carolina State University,
Raleigh, North Carolina 27695, USA}
\affiliation{Technische Universität Darmstadt, Department of Physics,
64289 Darmstadt, Germany}

\author{A.~Schwenk}
\email{schwenk@physik.tu-darmstadt.de}
\affiliation{Technische Universität Darmstadt, Department of Physics,
64289 Darmstadt, Germany}
\affiliation{ExtreMe Matter Institute EMMI and Helmholtz Forschungsakademie
  Hessen f\"ur FAIR (HFHF),
GSI Helmholtzzentrum für Schwerionenforschung GmbH,
64291 Darmstadt, Germany}
\affiliation{Max-Planck-Institut für Kernphysik, Saupfercheckweg 1,
69117 Heidelberg, Germany}

\begin{abstract}
We investigate the appearance of resonances in three-body systems
using pionless effective field theory at leading order
 with two complementary methods.
The Faddeev equation is analytically continued to
the unphysical sheet adjacent to the positive real energy axis using
a contour rotation. We consider both the three-boson system and the
three-neutron system. For the former, we calculate the trajectory of
Borromean three-body Efimov states turning into resonances as they
cross the three-body threshold. For the latter, we find no sign of
three-body resonances or virtual states at leading order.
This result is validated by exploring the level
structure of three-body states in a finite volume approach.
\end{abstract}

\maketitle

\section{Introduction}

The search for few-neutron resonances and bound states has a
long history with ambiguous results~\cite{Slaus:1972,Kezerashvili:2016ucn}.
In this work we focus on the topic of three-neutron resonances, motivated
by a controversial discussion of this topic in the recent
literature~\cite{Gandolfi:2016bth,Truol:2017yma,Deltuva:2018lug,%
Deltuva:2019ngx,Li:2019pmg,Ishikawa:2020bcs,Higgins:2020avy,Higgins:2020pbe}.

The first theoretical studies based on the analytical
continuation of the Faddeev equation in the 1970s using a
Yamaguchi-type two-neutron (\(nn\)) interaction in the \(^1S_0\) channel found
no evidence for a three-neutron resonance~\cite{Glockle:1978zz,Moeller:1979}.
Further experiments~\cite{Miller:1980pn,Stetz:1986zh,Tilley:1987svb,Yuly:1997ja}
and theoretical investigations~\cite{Offermann:1979wbx,Sofianos:1997nn} did not
satisfactorily resolve the situation.
Theoretical studies using the complex scaling method (CSM) in the 1990s
indicated a possible three-neutron resonance with an unphysically large
width~\cite{Csoto:1995ci,Witala:1999pm}.
These results were supported by Ref.~\cite{Hemmdan:2002tw}, extending Glöckle's
earlier work~\cite{Glockle:1978zz} to more 
partial-wave channels for the $nn$
interaction.
However, subsequent theoretical investigations based on the CSM and analytical
continuation in the coupling constant (ACC), again excluded a possible
three-neutron resonance~\cite{Lazauskas:2005fy}.

The interest in few-neutron resonances was revived in 2016, when experimental
evidence for a four-neutron resonance was presented
by~\textcite{Kisamori:2016jie}. A recent experiment even suggested
that the tetraneutron could be bound \cite{Faestermann:2022meh}.
A theoretical study of $3n$ and $4n$ systems suggested that these problems are
connected and a three-neutron
resonance might exist below a four-neutron resonance \cite{Gandolfi:2016bth},
which was subsequently supported by other work~\cite{Li:2019pmg}.
However, these results were criticized and led to a controversial
discussion~\cite{Truol:2017yma,Deltuva:2019ngx,Gandolfi:2019gmt}.
Further studies based on the Alt-Grassberger-Sandhas (AGS) equations for
transition operators~\cite{Deltuva:2018lug} and response
functions~\cite{Ishikawa:2020bcs} found no evidence of a three-neutron
resonance.
Higgins~\etal~\cite{Higgins:2020avy,Higgins:2020pbe} confirmed this further with
calculations in a hyperspherical framework.
They pointed out that there is significant attraction compared to free neutrons.
However, because of the Pauli repulsion it does not lead to a resonance but
shows up as a clear enhancement in the Wigner-Smith time delay.
Other recent studies investigated the spectral properties of
three-body systems near unitarity by mapping to Gaussian potentials
\cite{PhysRevC.102.064001}, and of nuclear systems with $A=3-6$, $16$
using two- and three-body contact interactions \cite{Schiavilla:2021dun}.
An overview of the theoretical and experimental situation regarding
few-neutron resonances was recently given in
Ref.~\cite{Marques:2021mqf}.

This overall situation is our motivation to investigate here the problem of
three-neutron resonances using pionless effective field theory (EFT)
\cite{vanKolck:1997ut,Kaplan:1998tg,Kaplan:1998we,vanKolck:1998bw}.
Because of the relevance for Efimov states in ultracold atomic gases
\cite{Braaten:2004rn,Naidon:2016dpf}, we also apply our method to three-boson
resonances.
Pionless EFT provides a controlled, model-independent description of few-body
systems with large two-body \(S\)-wave scattering length, based on an expansion
in the ratio of short- and long-distance scales (see Refs.~\cite{Beane:2000fx,
Bedaque:2002mn,Epelbaum:2008ga,Hammer:2019poc} for reviews).
This description breaks down for momenta of the order of the pion mass, but is
ideally suited to investigate the properties of low-energy neutron systems.
It allows for a model-independent assessment of the resonance question which can
be systematically improved by calculating higher orders.
In the scope of pionless EFT, all higher-order corrections,
including attractive $P$-wave channels, are perturbative. Thus we
do not expect higher-order corrections to alter the low-energy resonance
structure of the system. Finally, pionless EFT describes
the low-energy properties of neutron matter very well (see, e.g., Ref.~\cite{Schwenk:2005ka}) 
and is thus ideally suited for our purpose.

In addition to an EFT framework, we need a method to investigate the resonance
spectrum in this theory.
The various methods for studying few-body resonances can be separated into two
classes: approaches which perform calculations on the physical sheet for some
form of unphysical modification of the system (altered interaction strength,
adding an artificial trap), and approaches which perform calculations directly
on the unphysical sheet.
Both types of approaches have advantages as well as disadvantages.
While the calculation on the physical sheet is simpler, the analytical
continuation to the unphysical sheet can be delicate or even questionable.
Direct searches of resonance poles on the unphysical sheet are much more
complex, both numerically and conceptually, but generally lead to more robust
results.
This work uses a combination of both approaches, with a primary focus on direct
calculations.

For the latter, we use the well-known Faddeev formalism~\cite{Faddeev:1961},
formulated in momentum space.
Already in 1964, Lovelace proposed the method of contour rotation to analytically
continue the Faddeev equation to the unphysical sheets~\cite{Lovelace:1964}.
This formalism was extended independently to general integration contours by
Glöckle~\cite{Glockle:1978zz}, as well as by
Möller~\cite{Kuhn:1977hp,Moller:1977vy,Moller:1978,Moller:1978dm,Moeller:1981}.
Furthermore, these works introduced a modified equation structure
that is simpler to use.
The basic idea by Lovelace of a rotated contour was extended by Pearce and Afnan
and applied to several systems~\cite{Pearce:1984ca,Afnan:1991kb,Afnan:1993pb,%
Afnan:2015ahc,Gibson:2019occ}.
We apply this formalism in this work and note that it is
conceptually equivalent to the CSM~\cite{Aguilar.1971,Balslev.1971,%
Myo:2014ypa,Myo:2020rni,Lazauskas:2019ltg}, which performs a rotation in
coordinate space.\footnote{%
An alternative formulation of resonances is given by the Berggren basis,
which includes discretized resonance states explicitly in the completeness
relation~\cite{Berggren:1968zz,Berggren:1993zz}.}

We complement our direct searches for complex resonance energies by an
alternative approach that falls somewhere in between the two classes of methods
mentioned above.
Expanding upon early work for two-body
systems~\cite{Wiese:1988qy,Luscher:1991cf,Rummukainen:1995vs,Klos:2016fdb},
Ref.~\cite{Klos:2018sen} established that 
energy spectra in a periodic finite
volume can be used to identify few-body resonance states as avoided crossings of
energy levels as the size of the volume is varied.
This method is based on the Lüscher
formalism~\cite{Luscher:1985dn,Luscher:1986pf,Luscher:1990ux}, the key insight
of which is that the infinite volume \(S\) matrix governs the spectrum of a system
in finite volume.
An attractive feature of this method is that it does not require any contour
rotation or modification of the interaction (although the finite volume bears
some similarity to adding an artificial trap to confine the system).
It is therefore straightforward to apply, and Ref.~\cite{Klos:2018sen} developed
an efficient discrete variable representation (DVR) to numerically calculate
few-body systems in periodic boxes.
A drawback of the method is that currently only resonance energies can be
readily inferred from the finite-volume spectrum, while extracting information
about widths requires further formal work.
For our use of the method here, however, this is not a
concern.

This work is structured in the following way.
In Sec.~\ref{sec:eft}, we derive a Faddeev equation in partial-wave
basis for the three-boson and three-neutron system in pionless EFT at leading
order.
The equation is then analytically continued to the unphysical sheet
adjacent to the positive real axis in Sec.~\ref{sec:anacon},
using a rotation of the contour of integration.
Section~\ref{sec:results} applies this formalism to the three-boson and
three-neutron system.
For the three-boson system,
we calculate the resonance energy and width for a broad range of
negative scattering length and compare to results in the
literature.
We show that no three-neutron bound state is possible and calculate the pole
trajectory on the unphysical sheet for a bound two-neutron subsystem.
Searching for resonance poles and virtual states for an unbound subsystem up to
the physical \(nn\) scattering length, no indications for resonances or virtual
states in the $J^\pi=\frac{1}{2}^-$, $\frac{3}{2}^-$, and $\frac{1}{2}^+$ channels are found.
In Sec.~\ref{sec:FV}, we discuss the complementary finite-volume formalism to
extract resonance properties from avoided level crossings
in finite volume energy spectra. No evidence of avoided level crossings for the negative parity states is found, confirming the Faddeev results for the $J^\pi=\frac{1}{2}^-$ and $\frac{3}{2}^-$ channels.
Finally, a summary and outlook are given in Sec.~\ref{sec:conc}.

We emphasize that the novel aspect of our
work lies in the use of a model-independent pionless EFT approach, which is systematic and transparent. Moreover, our
study is carried out using two complementary methods with different systematics: 
resonance trajectories in the complex plane and finite-volume calculations.

\section{Faddeev formalism}
\label{sec:eft}

Since we work in the Faddeev formalism,
we follow Refs.~\cite{Platter:2004he,Platter:2004zs}
and use pionless EFT
\cite{vanKolck:1997ut,Kaplan:1998tg,Kaplan:1998we,vanKolck:1998bw}
to construct an effective interaction potential,
\begin{equation}
  {V}_\mathrm{eff} = \sum_{n=2}^\infty {V}_n \,,
\end{equation}
where the index $n$ specifies an $n$-body potential.
In general, interaction terms up to $n=N$ contribute in a $N$-body
problem, but at low energies higher-body terms are
typically suppressed. The potentials ${V}_n$
are constrained by Galilean invariance and thus depend only on
the relative momenta.
They can be expressed in a momentum expansion, e.g.,
\begin{equation}\label{eq:2BInt}
  \langle \vec{k'}|{V}_2|\vec{k}\rangle =
  C_0 + C_{2}(\vec{k'}^2 + \vec{k}^2)/2 +\ldots
\end{equation}
for \( S \)-wave two-body interactions,
where $\vec{k}$ and $\vec{k'}$ are the relative momenta in the
initial and final states and regulator functions have been suppressed.
Similar expressions can be derived
for three- and higher-body interactions. At leading order in pionless EFT,
only a momentum-independent two-body contact interaction in the $^1S_0$
channel contributes for the three-neutron system
\cite{Bedaque:1997qi,Bedaque:1998mb}.
In the three-boson system, in contrast, an additional momentum-independent 
three-body contact interaction has to be included to properly
renormalize the system \cite{Bedaque:1998kg,Bedaque:1998km}.
Assuming typical momenta of order $1/a$, the uncertainty of a leading-order
pionless EFT calculation can be estimated as $|r/a|$, where $r$ is the effective
range and $a$ the scattering length. For the three-neutron system, we have
$a\approx -18.9$ fm and
$r\approx \SI{2.7}{\femto\meter}$~\cite{Gardestig:2009ya},
such that a leading-order calculation has an
uncertainty of about 15\%. Thus the higher-order corrections are not
expected to change the resonance structure of the system.
The exact form of the
effective potential depends on the specific regularization scheme used.
The low-energy observables, however, are independent of the regularization
scheme (up to higher-order corrections) and one can choose a convenient
scheme for practical calculations. Explicit forms for the effective
potentials will be given below. (For a more detailed discussion of pionless EFT
including a more formal discussion of the power counting, we refer to the
reviews~\cite{Beane:2000fx,Bedaque:2002mn,Epelbaum:2008ga,Hammer:2019poc}.)

Our starting point for deriving the Faddeev equations is the
full relative
three-body wave function \(\ket{\Psi}\), defined as a solution of
the stationary Schr\"odinger equation.
This wave function is decomposed into the three so-called Faddeev components
\(\ket{\psi_i}\) according to
\begin{equation}
 \ket{\Psi} = \sum_{i=1}^3 \ket{\psi_i} \equiv G_0 \left( \sum_{i=1}^3
 V_2^{(i)} + V_3\right) \ket{\Psi} \,,
\end{equation}
where $i=1,2,3$ labels the three different pairs in the three-body system.
The above definition includes the free Green's function
\begin{equation}
 G_0(z)  = \frac{1}{z-H_0} \,,
\end{equation}
where \(z\) is an arbitrary (in principle complex) energy.
\(H_0\) represents the relative kinetic part of the Hamiltonian and the two-body
pair interactions are given by \(V_2^{(i)}\).
Moreover, a three-body force \(V_3\) is included here
as well
to keep the
equation sufficiently general for the three-boson and
three-neutron systems.

Introducing the permutation operator
\begin{equation}
 P = P_{12}P_{23} + P_{13}P_{23}\,,
\end{equation}
we are able to express the full state by only one component,
\begin{equation}
 \ket{\Psi} = \left(1+P\right) \ket{\psi_1}.
\end{equation}
The index 1 is dropped in the following.

Altogether, the leading-order representation of the Faddeev equation
is given by~\cite{Stadler:1991zz}
\begin{equation}\label{eq:Faddeev}
 \ket{\psi} = G_0 t P \ket{\psi} + \frac{1}{3} \left( G_0 + G_0 t G_0 \right)
 V_3 \left(1+P\right) \ket{\psi}.
\end{equation}
The \(S\)-wave two-body interaction \(V_2\) is chosen as
\begin{equation}
  \matrixel{k'}{V_2}{k} =C_0 \braket{k'}{g} \braket{g}{k}\,,
\end{equation}
with strength \(C_0\), $k=|\vec{k}|$, and $k'=|\vec{k'}|$.
It is included via the two-body
\(T\) matrix \(t\) which satisfies the Lippmann-Schwinger equation.
We use a Gaussian type regulator function
\begin{equation}
 \braket{p}{g} = g(p) = \exp\left({-}p^2/\Lambda^2\right),
\end{equation}
where \(\Lambda\) is the cutoff scale.
This form is particularly convenient for the analytic continuation of the
formalism into the complex plane.
For the finite-volume calculations discussed in Sec.~\ref{sec:FV}, we will
also consider super-Gaussian regulators, which fall off faster at large
momenta, to improve the convergence.

The three-body potential is parametrized as
\begin{equation}\label{eq:TBInt}
 V_3 = D_0 \ket{\xi}\bra{\xi},
\end{equation}
with the interaction strength \(D_0\) and the three-body regulator
\(\ket{\xi}\).
We again choose a (separable) Gaussian regulator function, connected to
the two-body regulator by
\begin{equation}\label{eq:SepGaussian}
\braket{u_1 u_2 }{\xi} = \xi(u_1,u_2) =
g(u_1) g\left(\frac{\sqrt{3}}{2}u_2\right),
\end{equation}
where
\begin{equation}
 \begin{aligned}
  \vec{u}_1 &= \frac{1}{2}\left( \vec{k}_1 - \vec{k}_2\right), \\
  \vec{u}_2 &= \frac{2}{3}
\left[ \vec{k}_3 - \frac{1}{2}\left( \vec{k_1}+\vec{k}_2\right)\right],
 \end{aligned}
\end{equation}
are three-body Jacobi momenta.
Here, \(\vec{u}_1\) represents the relative momentum between the first two
particles, while the relative momentum between the third particle and the
center of mass of the first two particles is given by \(\vec{u}_2\).
The relative kinetic energy of the three-body system is given by
\begin{equation}
H_0 \ket{\vec{u}_1\vec{u}_2}
= \left(u_1^2 + \frac{3}{4}u_2^2 \right)\ket{\vec{u}_1\vec{u}_2}.
\end{equation}
Here and in the following, we set \(m=1\) such that energy and momentum
squared have the same units.
Together with the appropriate angular momentum, spin, and isospin quantum
numbers, which are summarized in the multi-index \(\ket{\mathrm{i}}\), we define
our basis as \(\ket{u_1 u_2 \mathrm{i}}\).

This work uses a \(jj\) coupling scheme, for which the set of quantum numbers is
given by
\begin{equation}
 \ket{\mathrm{i}} = \ket{(ls)j (\lambda s_3)I J},
\end{equation}
with the relative orbital angular momentum \(l\) between the first two particles
and the orbital angular momentum \(\lambda\) relative to the third particle.
\(s\) is the coupled spin of the first two particles, which couples with \(l\)
to \(j\).
Similarly, the spin of the third particle \(s_3\) couples with \(\lambda\) to
\(I\), which itself is coupled with \(j\) to the total angular momentum~\(J\).

We now derive the equations for both the three-neutron and three-boson
systems in parallel.
For definiteness, we consider a system of three spinless bosons, where
\begin{equation}
 \ket{\mathrm{i}} = \ket{(00)0 (0 0)0 0}.
\end{equation}
In the case of the three-neutron system, we suppress the isospin quantum
number $3/2$, while the three-body force is
absent because the Pauli principle precludes a momentum-independent
contact three-neutron force.\footnote{Three-neutron forces including derivatives
would be permitted, but they only enter at higher orders in the EFT power counting.}
To leading order in pionless EFT, only \(S\)-wave two-body
interactions contribute.
We consider the basis states
\begin{equation}
 \ket{\mathrm{i}} = \ket{(00)0 (1 \frac12)\frac32 \frac32}\,,
\end{equation}
which correspond to $J^\pi= \frac{3}{2}^-$, and
\begin{equation}
 \ket{\mathrm{i}} = \ket{(00)0 (\lambda \frac12)\frac12 \frac12}\,,
\end{equation}
for the cases $\lambda=0,1$,
which correspond to $J^\pi = \frac{1}{2}^+$, $\frac{1}{2}^-$.
As will be shown below, the Faddeev equations for \(\lambda=1\) are the same
and will be investigated simultaneously.
In all cases, the Faddeev equations reduce to a single
channel, and consequently the index \(\mathrm{i}\) will be dropped in the
following.

Now we are able to derive the three-body partial-wave projected Faddeev equation
by projecting \eqref{eq:Faddeev} onto the single-channel basis.
We exploit the fact that the free Green's function \(G_0\) is diagonal in all variables,
\begin{equation}
 \begin{aligned}
 &\matrixel{u_1^\prime u_2^\prime}{G_0(E)}{{u}_1 {u}_2} \\
 &\hspace*{1cm}= G_0\left(E; {u}_1, {u}_2 \right)
 \frac{\delta\left( u_1^\prime-{u}_1 \right)}{u_1^\prime {u}_1}
 \frac{\delta\left( u_2^\prime-{u}_2 \right)}{u_2^\prime {u}_2}\,,
\end{aligned}
\end{equation}
with
\begin{equation}
  \label{eq:g0}
 G_0(E;u_1,u_2) = \left( E - u_1^2 - \frac{3}{4}u_2^2 \right)^{-1}.
\end{equation}
The two-body \(t\)-matrix can be written as
\begin{equation}
 \begin{aligned}
  &\matrixel{u_1^\prime u_2^\prime}{t(E)}{u_1 u_2} \\
  &\hspace*{1cm}= g(u_1^\prime) \tau\left(z\right)
  g(u_1)
  ~\frac{\delta(u_2^\prime-u_2)}{u_2^\prime u_2}\,,
 \end{aligned}
\end{equation}
with the energy of the first pair of particles
\(z=E-\frac{3}{4}u_2^{\prime 2}\).
Within the EFT formalism \(\tau\) describes the propagation of an
interacting two-particle state, commonly called a ``dimer.''
Following this convention, we refer to \(\tau\) as
the dimer propagator.
Together with the explicit representation of the three-body
interaction~\eqref{eq:TBInt}, the Faddeev equation can be written as
\begin{widetext}
\begin{equation}\label{eq:FadFinal}
 \begin{aligned}
  \psi \left({u}_1 {u}_2 \right) &=
  ~G_0\left(E; {u}_1, {u}_2 \right) \Bigg[  \int_{-1}^{+1} \mathrm{d}x
  \int \mathrm{d}u_2^{\prime} u_2^{\prime 2}  ~g(u_1)
  \tau\left(E-\frac{3}{4}u_2^{2}\right) g(\pi_1)
  G(u_2 u_2^{\prime}x)
  \braket{\pi_2u_2^{\prime}}{\psi}   \\
  &\qquad\qquad+ D_0   \int \mathrm{d}u_1^{\prime} u_1^{\prime 2}
  \int \mathrm{d}u_2^{\prime} u_2^{\prime 2}~\xi(u_1^\prime, u_2^\prime)
  \braket{u_1^\prime u_2^\prime}{\psi}
  \Bigg\{ \xi(u_1, u_2) +
  g(u_1)   \tau \left(E-\frac{3}{4}u_2^{2}\right)  \\
  &\hspace*{2.8cm}\times
  \int \mathrm{d}u_1^{\prime\prime} u_1^{\prime\prime 2}
  ~\xi(u_1^{\prime\prime},u_2) G_0\left(E;{u}_1^{\prime\prime},
  {u}_2 \right)
  g(u_1^{\prime\prime})\Bigg\}\Bigg].
 \end{aligned}
\end{equation}
\end{widetext}
Note that the factor \(1+P\) within Eq.~\eqref{eq:Faddeev} cancels against the
factor \(1/3\).
Here we have used the matrix element of the permutation operator
\begin{equation}
 \begin{aligned}
  &\matrixel{u_1^\prime u_2}{P}{u_1^{\prime\prime}u_2^{\prime\prime}} \\
  &\hspace*{0.65cm}= \int_{-1}^{+1} \mathrm{d}x
  \frac{\delta(u_1^\prime-\pi_1)}{u_1^{\prime 2}}
  \frac{\delta(u_1^{\prime\prime}-\pi_2)}{u_1^{\prime \prime
  2}} G (u_2 u_2^{\prime\prime}x)\,,
 \end{aligned}
\end{equation}
with
\begin{align}
 \pi_1 &= \sqrt{u_2^{\prime\prime 2} + \frac{1}{4} u_2^{2}
 + u_2u_2^{\prime\prime}x}\,, \\
 \pi_2 &= \sqrt{\frac{1}{4} u_2^{\prime\prime 2} + u_2^{2}
 + u_2u_2^{\prime\prime}x}\,.
\end{align}
In general, the recoupling function \(G (u_2 u_2^{\prime\prime}x)\)
depends on both momenta and angular quantum numbers.
Besides the two-body \(t\)-matrix, it is this term that
mainly incorporates the information about the quantum numbers of the system.
It reduces to the Legendre polynomial $P_0$ for the three-boson system,
\begin{equation}\label{eq:G3b}
  \begin{aligned}
    G_\mathrm{3b} (u_2 u_2^{\prime\prime}x) = P_0(x) = 1\,,
  \end{aligned}
\end{equation}
and to a constant times a Legendre polynomial for the three-neutron system,
\begin{equation}\label{eq:G3n}
  \begin{aligned}
   G_\mathrm{3n}^\lambda (u_2 u_2^{\prime\prime}x)
   = {-}\frac{1}{2} P_\lambda(x)\,,
  \end{aligned}
\end{equation}
for $\lambda=0,1$.
Finally, the dimer propagator \(\tau\) can be written as
\begin{equation}
 \begin{aligned}
  \tau(z) &= \left[ \frac{1}{C_0} - I(z)
   \right]^{-1}, \qquad\mbox{with}\\
  I(z) &=  \matrixel{g}{G_0(z)}{g} =
\int_0^\infty \mathrm{d}q ~q^2 \frac{g(q) g(q)}{z+i\varepsilon-q^2}\,.
 \end{aligned}
\end{equation}
Solving the integral \(I(z)\) analytically results in
\begin{widetext}
\begin{align}\label{eq:Tau}
 \begin{aligned}
  \tau(z) &= \frac{2}{\pi}\Bigg[ \gamma \exp \left(2
  \frac{\gamma^2}{\Lambda^2} \right) \mathrm{erfc}\left(
  \frac{\sqrt{2}|\gamma|}{\Lambda} \right)  +  i \sqrt{z}
  \exp \left( -2 \frac{z^2}{\Lambda^2} \right)
  \mathrm{erfc}\left( \mp i \frac{\sqrt{2z}}{\Lambda} \right)
  \Bigg]^{-1} \\
  &=\frac{2}{\pi} \Bigg[
  \gamma + i \sqrt{z}
  + \mathcal{O}\left(\Lambda^{-1}\right)
  \Bigg]^{-1}.
 \end{aligned}
\end{align}
\end{widetext}
The upper (lower) sign corresponds to the case \(\Im z > 0\) (\(\Im z < 0\)),
while \(\mathrm{erfc(z)}\) represents the complementary error function,
\begin{equation}
 \mathrm{erfc}(z) = 1 - \erf(z)
 = \frac{2}{\sqrt{\pi}} \int_z^\infty \mathrm{d}t ~e^{-t^2}.
\end{equation}
The representation in the first line of Eq.~\eqref{eq:Tau} includes
finite-range contributions induced by the finite cutoff $\Lambda$.
The expressions in the first and second line are equivalent in the limit
\(\Lambda \longrightarrow \infty\).
Within this work we use the representation in the second line because it
provides a simpler analytic continuation to complex resonance energies.
We renormalize the dimer propagator by choosing \(C_\mathrm{0}\)
to reproduce a pole in the two-body subsystem
at \(\sqrt{z}=i\gamma\).
For positive \(\gamma\), we reproduce a two-body bound
state at \(E_2 = - \gamma^2 = -1/a^2\), with the two-body binding
momentum \(\gamma\) and the scattering length \(a\).
For negative  \(\gamma\), we reproduce the corresponding virtual
state.

Finally, since we are working with separable interactions, it is convenient to
transform the Faddeev equations by defining
\begin{equation}\label{eq:RedFadComp}
 \psi(u_1, u_2) = G_0(E;u_1, u_2) g(u_1)
 \tau\left(E-\frac{3}{4}u_2^2\right) F(u_2)\,,
\end{equation}
where \(F\left(u_2\right)\) is the so-called reduced Faddeev component.
Instead of the full Faddeev component the reduced component only
depends on one momentum variable reducing the numerical effort to solve
the problem.

\subsection{Three-boson equation}

Combining all contributions presented above, the Faddeev equation representing
the three-boson system is given by
\begin{equation}\label{eq:TBfinal}
 \begin{aligned}
 F(u_2)&= \int \mathrm{d}u_2^\prime ~u_2^{\prime 2}\,
 \tau\left(z\right)
 \left(Z_2 + Z_3 \right) F(u_2^\prime)\,.
 \end{aligned}
\end{equation}
In analogy to the Lippmann-Schwinger equation, we define a
two-body interaction kernel
\begin{equation}\label{eq:Z2}
 Z_2 = \int_{-1}^{+1} \mathrm{d}x ~g(\pi_1)G_0(E;\pi_2, u_2^\prime) g(\pi_2)
\end{equation}
with \(z=E-\frac{3}{4}u_2^{\prime 2}\) and three-body interaction kernel
\begin{equation}
 Z_3 = \frac{D_0}{C_0}
 g\left(\frac{\sqrt{3}}{2}u_2\right)
 g\left(\frac{\sqrt{3}}{2}u_2^\prime\right)
 I\left(E-\frac{3}{4}u_2^{\prime 2}\right).
\end{equation}
While \(Z_3\) arises from the three-body force,
\(Z_2\) is the contribution from the one-particle exchange.

\subsection{Three-neutron equation}

Finally, we adapt the equation to the three-neutron system.
Due to the Pauli principle only an \(S\)-wave
$nn$ interaction in the \(^1S_0\) channel is possible.
The exact value of the $nn$ scattering length is still debated, but
the currently accepted  value is
$(-18.9\pm 0.4)$~fm~\cite{Gardestig:2009ya}.
The third neutron is to be assumed in a relative
\(P\) wave (\(\lambda=1\)) in accordance with previous studies.
This results in the possible states \(J^\pi = \frac{1}{2}^-\) and
\(\frac{3}{2}^-\), which are degenerate in leading-order pionless EFT. These are the most likely channels for three-neutron resonances to occur \cite{Marques:2021mqf}. The corresponding Faddeev equation reads
\begin{equation}\label{eq:3n-Fad}
 \begin{aligned}
  F(u_2) &= - \frac{1}{2} \int \mathrm{d}u_2^\prime u_2^{\prime 2}
   \int_{-1}^{+1}\mathrm{d}x ~g(\pi_1) G_0(E;\pi_2, u_2^\prime)
   \\
   &\qquad \times g(\pi_2) P_1(x)
   \tau\left(E-\frac{3}{4}u_2^{\prime 2}\right) F(u_2^\prime)\,.
 \end{aligned}
\end{equation}
Moreover, we consider the case of the third neutron in a relative \(S\)-wave
(\(\lambda=0\)), corresponding to \(J^\pi = \frac{1}{2}^+\). The resulting Faddeev equation is derived from Eq.~\eqref{eq:3n-Fad}
by replacing \(P_1(x)\) by \(P_0(x)=1\).
In contrast to the three-boson system, the three-neutron equations feature no three-body forces since such terms are highly suppressed for identical fermions.
This is taken into account by setting \(D_0\) to zero.

\section{Analytical Continuation: Method}
\label{sec:anacon}

Due to the square root connection between the energy and the momentum variables,
two points in the complex momentum plane are mapped onto a single point in the
complex energy plane.
This mapping is made unique by introducing a two-sheet structure  for the energy
variable.
Solving the Faddeev equations for the three-body system, it is straightforward
to search for bound states located on the first (or ``physical'') sheet of the
complex energy plane.
In this work, however, we are interested in resonances and virtual states, which
live on the second (``unphysical'') sheet.

The procedure described in the following assumes a three-body system of
identical particles, for which the two-body subsystems are not bound.
It can easily be extended to bound subsystems by considering the complex energy
and momentum planes relative to the two-body threshold.
We note that the formalism can also be applied to systems of nonidentical
particles.
This leads to a more complicated sheet structure due to further thresholds.
In this work, however, we need not deal with this complication.

\begin{figure*}[htp]
  \includegraphics{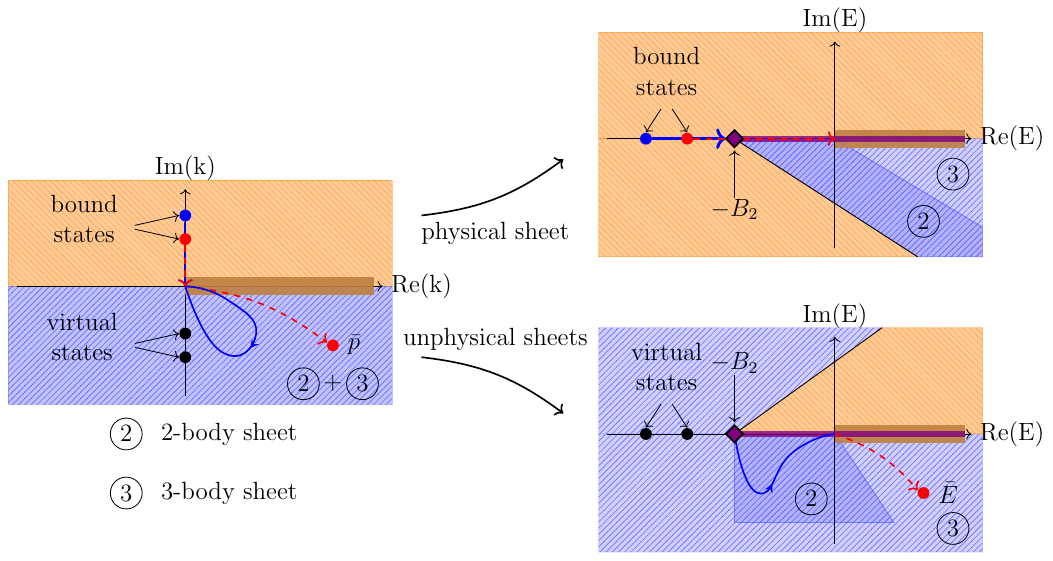}
\caption{
The structure of the complex momentum and energy plane for three particles
defined by the energy-momentum relation \(E= \frac{3}{4}k^2 - B_2\), where
\(B_2\) is the two-body binding energy.
Energies on the physical sheet (upper right plot) translate to momenta on the
physical part of the momentum plane with \(\Im(k) >0\)
(shading \shadingbackslash).
Energies on the unphysical sheets (lower right plot) are mapped to the region
of the complex momentum plane with \(\Im(k)<0\)
(shading \shadingslash).
The physical and unphysical sheets are connected by two branch cuts;
the three-body cut starting at the origin and following the
positive real axis and the two-body cut starting at the two-body
binding energy \(B_2\) (\tikzdiamond \!) and following the real axis, too.
The complex energy plane shows two unphysical sheets: the one accessible
through the cut starting at the two-body threshold (2-body sheet, darker shaded)
and the one accessible through the three-body threshold (3-body sheet,
lighter shaded).
Both unphysical sheets extend further than sketched here.
While bound states are located on the physical sheet, virtual states and
resonances \(\bar{p}/ \bar{E}\) live on the unphysical sheets.
This is also true for the corresponding areas on the complex momentum plane.
The pole trajectories as a function of the two-body interaction strength
of the three-boson system (\(a<0\),
dashed lines) and three-neutron
system using the Yamaguchi model (solid lines) are sketched.
Starting at a given bound state and decreasing the two-body
interaction strength the three-boson system moves through the three-body branch
point, as the two-boson system is unbound, and evolves into a resonance.
The Yamaguchi model allows to create a bound two-neutron subsystem, of which
the binding energy decreases slower than the three-neutron binding energy.
At some point both values are equivalent and the pole trajectory moves through
the two-body cut onto the unphysical sheet.
Finally, the two as well as the three-neutron pole trajectory meet again at the
origin.
Note that the position of the two-body branch point depends on the two-body
interaction strength. So its position is different for every point along the
pole trajectory.}
\label{fig:ComplexMomPlane}
\end{figure*}

\medskip
We start at a three-body bound state \(E^{(0)}\) for an (unphysical) scattering
length \(a^{(0)}>0\) and investigate the pole trajectory in the complex momentum
plane as a function of the scattering length.
In systems of ultracold atoms, these trajectories can be followed
experimentally using Feshbach resonances~\cite{RevModPhys.82.1225}.
\Cref{fig:ComplexMomPlane} shows a sketch of such a trajectory in the
relevant region of the complex momentum plane.

A characteristic point of this trajectory is the origin, which  corresponds to a
so-called branch point.
Two sheets are connected by a branch cut, which is spanned between two branch
points.
The first method used in this paper to search for resonances is to analytically
continue the Faddeev equations derived in the previous chapter through this cut
onto the adjacent unphysical sheet.

In the Faddeev equations a cut can originate from either of two characteristic
structures in the kernel:
On the one hand, there is the dimer propagator, Eq.~\eqref{eq:Tau}.
This square-root branch cut is not relevant for this work as it is only present
for bound two-body subsystems.
On the other hand, the kernel includes the one-particle exchange contribution
given in Eq.~\eqref{eq:Z2}.
The relevant structure is the free Green's function
\begin{equation}\label{eq:PotG0}
 G_0(E;\pi_2, u_2^\prime) = \Big[ E - u_2^2 - u_2^{\prime 2}
 -u_2 u_2^\prime x \Big]^{-1} .
\end{equation}
It generates a branch cut, the so-called three-body cut, between the branch
point at the origin (\(u_2 = u_2^\prime = 0\)) and the second branch point at
infinity in the limit \(u_2, u_2^\prime \longrightarrow \infty\).
Applying the partial-wave projection by integrating over \(x\)  results in a
logarithmic structure.
Similarly to the square root, the complex logarithm is a multivalued function: it
does not change if an integer multiple of \(2\pi i\) is added to its argument.
Therefore, this branch cut leads to an infinite number of unphysical sheets.
Physically, only the one adjacent to the lower rim of the physical sheet is
relevant as it affects measurable quantities such as the cross
section.\footnote{Note that the regulator can generate also an artificial cut.
However, the contact interaction with separable Gaussian regulator used here
does not generate any singularities.}

\begin{figure}[t]
  \includegraphics{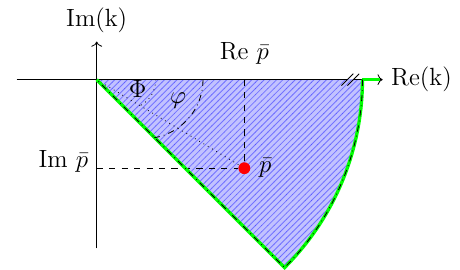}
  \caption{The part of the unphysical region adjacent to the three-body cut,
  which is available if the angle of rotation is chosen to be \(\varphi\).
  The angle has to be larger than the angle \(\Phi\) of a possible state
  \(\bar{p}\) on that sheet.}
  \label{fig:Afnan}
 \end{figure}

Assuming the pole moves through the cut onto the physically relevant unphysical
sheet, we have to analytically continue the Faddeev equations to momenta with a
negative imaginary part and a positive real part, \ie, to complex momenta in the
lower right quadrant of the complex plane.
The analytical continuation is based on the general idea of writing down a
Faddeev-like equation with a contour of integration on the unphysical sheet.
This procedure moves the location of the cut and thereby makes part of the
second sheet accessible via the standard Faddeev equation, \ie, an equation that
is formally the same as before except that the integral runs along the contour
in the complex plane.
Using this equation, all poles, which are located between the rotated contour
(which coincides with the rotated branch cut) and the positive real axis can be
identified.
Generally, there are an infinite number of possible contours.
However, it is sufficient in practice (and convenient) to use just one
particular type of contour, which was first applied to this problem by Pearce
and Afnan~\cite{Pearce:1984ca}.
The contour is constructed by rotating the integral from the positive real axis,
$[0,\infty)$, into the lower right quadrant,
\begin{equation}\label{eq:MomRot}
 u_2^\prime \longrightarrow u_2^\prime ~e^{-i\varphi}, \qquad \varphi >0.
\end{equation}
The benefit of this contour is that it is rather simple and characterized by
only one parameter, the rotation angle \(\varphi\).\footnote{In principle this
 contour should be constructed by rotating a finite interval (say,
 $[0,\Lambda]$) and then closing it towards the real axis at the end point.
 However, we assume here that the contribution from the
 arc becomes irrelevant in the limit $\Lambda\to\infty$.}
This angle has to be chosen such that the cut is rotated beyond the position of
the state of interest (cf.~\cref{fig:Afnan}).
This statement is equivalent to the condition
\begin{equation}
 \varphi > \Phi = \arctan \Bigg| \frac{\Im \bar{p}}{\Re \bar{p}} \Bigg|.
\end{equation}

The key requirement of this formalism is that the kernel of the Faddeev equation
is analytic within the momentum region covered by the contour rotation.
First, we have to show that these type of contours can be used without
encountering a singularity.
As explained before, we here need only take into account poles of the
one-particle exchange contribution.
Considering the momentum plane of the integration variable \(u_2^\prime\),
we choose momenta \(u_2\) along the contour of integration for a fixed energy
\(E\).
This results in two areas in which the kernel is not analytic in~\citep{Glockle:1978zz}.
The general procedure is to start with an energy on the physical sheet and show
that the contour of integration can be deformed in such a way that eventually
the interesting part of the unphysical sheet becomes accessible without hitting
any one of the singular areas.
This process takes place in repeated steps.
First, the contour is rotated as far as possible.
Then, the energy is moved as far as possible, too.
These steps are repeated as often as necessary.

Further, we have to require that the momentum of the first pair of particles
\(k\) is continuous within the momentum plane.
Again, we start at a bound state \(E^{(0)} < 0\) on the physical sheet.
So, we apply the three-boson or three-neutron Faddeev equation with the
integration contour along the positive real axis (\(u_2^{\prime 2} \in
\mathbbm{R}^+\)).
Here, \(k\) is defined by
\begin{equation}
 \begin{aligned}
  k = \sqrt{z} &= \sqrt{E - \frac{3}{4} u_2^{\prime 2} + i\varepsilon}\\
       &= i \sqrt{-E + \frac{3}{4} u_2^{\prime 2}  - i\varepsilon}\,.
 \end{aligned}
\end{equation}
Both representations of the square root are equivalent and the
\(i\varepsilon\) term is explicitly needed to determine the correct
branch.

Now, we move to the unphysical sheet of interest with
\(E, u_2^{\prime 2} \in \mathbbm{C}\) and \(\Im E, \Im u_2^\prime <0 \).
Here, the \(i\varepsilon\) is not necessary anymore and we drop it.
So, also the two representations are no longer equivalent.
Choosing \(u_2^\prime\) along the contour of integration and \(E\) within the
fourth quadrant of the complex energy plane where the resonances live, we have
to check that \(k\) does not cross any cut.
This is only the case for one of the two representations.
Choosing \(\varphi=\Phi\) the argument of both representations crosses the real
axis at the origin.
As mentioned above, we choose \(\varphi>\Phi\).
So, the argument of the first representation crosses the real axis at positive
real parts, while the second representation crosses the axis at negative
imaginary parts.
The complex square root is a two-branched function.
Both branches are connect at the negative real axis.
The second representation would cross the square-root branch cut, while the
first does not.
So, for energies in the fourth quadrant we have to use the first representation.
A similar investigation shows that also for energies in the third quadrant the
first representation is the one to use.
Note that these statements are only correct for an unbound two-body subsystem.
If the subsystem is bound we have to consider the energy plane which is shifted
by the two-body binding energy.
This results in a dependence of the rotation angle on the two-body binding
energy.

\medskip
Altogether, the analytically continued Faddeev equation for the three-boson
system is given by Eq.~\eqref{eq:TBfinal} with \(\tau\) given by
Eq.~\eqref{eq:Tau} and the contour rotation Eq.~\eqref{eq:MomRot},
\begin{equation}
 \begin{aligned}
  F(u_2e^{-i\varphi})
  &= \int \mathrm{d}u_2^\prime ~u_2^{\prime 2} e^{-3i\varphi}\,
  \tau\left(z\right) \\
  &\qquad \times
  \left(Z_2 + Z_3 \right) F\left(u_2^\prime e^{-i\varphi}\right)\,.
 \end{aligned}
\end{equation}
Applying the same modifications to Eq.~\eqref{eq:3n-Fad}, the
analytically continued three-neutron equation for $\lambda=1$
is given by
\begin{equation}\label{eq:3nContinued}
 \begin{aligned}
  F(u_2&e^{-i\varphi}) \\
  &= - \frac{1}{2} \int \mathrm{d}u_2^\prime ~u_2^{\prime 2}
  e^{-3i\varphi} F(u_2^\prime e^{-i\varphi})\\
  &\qquad  \times
   \int_{-1}^{+1}\mathrm{d}x ~g(\pi_1 e^{-i\varphi}) g(\pi_2 e^{-i\varphi})  P_1(x)
   \\
   &\qquad  \quad \times G_0(E;\pi_2 e^{-i\varphi}, u_2^\prime e^{-i\varphi})
   \tau\left(z\right) \,.
 \end{aligned}
\end{equation}
The corresponding equation for $\lambda=0$ is obtained by
substituting $P_1(x) \to 1$ in Eq.~(\ref{eq:3nContinued}).

\medskip
Finally, let us add a remark on the solution of the Faddeev equations.
The mathematical structure of the inhomogeneous Faddeev equation
for the \(T\) matrix is given by a so-called Fredholm equation of the
second kind.
On the physical sheet the kernel is Hermitian and it can be shown that
the \(T\) matrix can be expanded in a basis given by the reduced
Faddeev components.
The coefficients of the expansion are proportional to 1 over
\(1-\lambda_n\), with the eigenvalues of the homogeneous
Faddeev equation \(\lambda_n\).
This expansion was extended to non-Hermitian kernels by
Afnan~\cite{Afnan:1991kb}.
So, to find poles on the unphysical sheet we have to search for eigenvalues
equal to one of the Faddeev equations for the reduced Faddeev components
along the rotated contour.
The first step of the procedure is now similar to the search for a bound
state on the physical sheet.
We expand the kernel in a momentum space basis derived by a
Gauss-Legendre mesh.
Now, these momenta are substituted by the rotated momenta.
Here, also the weight in the integral has to be transformed correctly.
Following the expansion, the next step would be to search for eigenvalues
equal to one as a function of the complex energy.
However, mathematically equivalent but numerically easier and faster
is the search for zeros of the characteristic polynomial for eigenvalues equal
to 1.
In comparison to the search for a bound-state pole this corresponds to
two-dimensional root finding.
Mathematically, this is much more advanced and it cannot be guaranteed
that the corresponding numerical routines will find all poles.
So, before applying the root-finding routines it is recommended to plot the
absolute value of the characteristic polynomial as a function of the complex
energy.
This plot largely depends on the numerical parameters used within the
derivation of the kernel matrix.
The only physically relevant part are the zeros, which are used as starting
values for the root-finding routines.

\FloatBarrier

\section{Analytical Continuation: Results}\label{sec:results}

We now discuss our results from the analytical continuation, first
for the three-boson case, where we can compare to previous studies
of resonances to benchmark our method, and then for the three-neutron case.

\subsection{Three-boson system}

\begin{figure}[t]
 \includegraphics{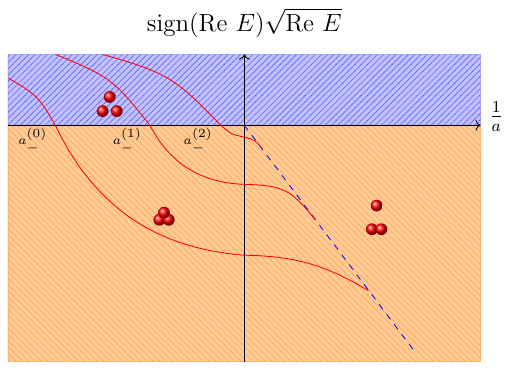}
 \caption{The so-called Efimov plot presenting the Efimov effect.
 The \(x\) axis presents the inverse scattering length, while the \(y\) axis
 shows the square root of the real part of the energy multiplied by
 its sign.
 The dashed-blue line indicates the binding energy of the dimer
 \(B_2 = 1/a^2\).
 In the area right to the dashed blue line the three-body system
 is unbound, while the dimer is bound.
 Crossing the line, the three-body system becomes bound too until a
 negative scattering length \(a_-^{(n)}\) is reached.
 Here, the pole moves from the physical sheet
 (light shaded) to the unphysical sheet adjacent
  the positive real axis (dark shaded) becoming a resonance.}
 \label{fig:Efimov}
\end{figure}

The three-boson system for large scattering length exhibits the so-called Efimov
effect~\cite{Efimov:1970zz,Braaten:2004rn}.
It leads to a universal spectrum of three-body bound states
which is illustrated in \cref{fig:Efimov}.
There is a more general discrete scaling symmetry which relates
the trajectory of any three-body bound state to all other states
via the transformation
\begin{equation}
 \begin{aligned}
  a &\longrightarrow \nu^n a\,,\\
  E &\longrightarrow \nu^{-2n} E\,,
 \end{aligned}
\end{equation}
where $n$ is an integer and $\nu \approx 22.7$ is the discrete
scaling factor.
In the unitary limit (\(a\longrightarrow \pm\infty\))
the binding energies of two consecutive states are connected by
\begin{equation}
 E_{n+1} = \nu^2 E_n\,.
\end{equation}
Here, the counting starts at the deepest bound state accessible within the EFT.
These discrete scaling symmetries are evident in \cref{fig:Efimov}.
Similarly, it is possible to connect the scattering lengths at which the pole
trajectory moves from the physical to the unphysical sheet by
\begin{equation}
 a_-^{(n)} = \nu a_-^{(n+1)}.
\end{equation}

Further, the symmetry manifests itself in a log-periodic behavior of the
three-body observables.
For convenience we introduce a dimensionless coupling \(H(\Lambda)
=D_0/ \Lambda^4\).
We renormalize \(H(\Lambda)\) at \(\gamma= 0\)
such that the energy of the shallowest three-body bound state keeps fixed
when varying the regulator scale \(\Lambda\).
This work uses two different renormalization prescriptions for \(H(\Lambda)\).
On the one hand, we choose a natural value for \(\Lambda\) and determine
\(H(\Lambda)\) to reproduce some three-body observable.
On the other hand, we choose \(\Lambda\) such that \(H(\Lambda)=0\)
\cite{Hammer:2000nf}.

\begin{figure}[htp]
  \includegraphics[width=0.5\textwidth]{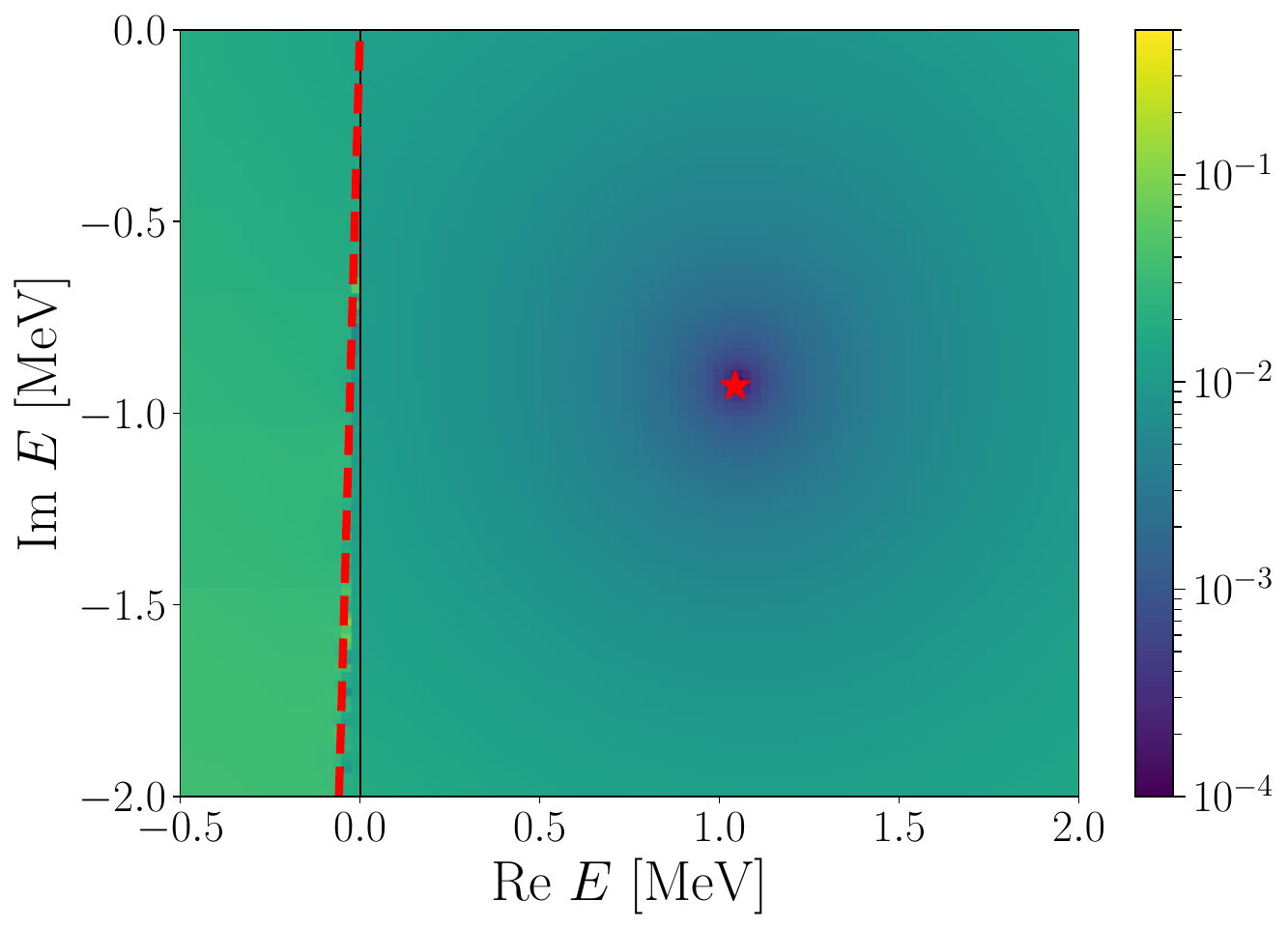}
  \caption{The absolute value of the characteristic polynomial for a scattering
    length \(a=\SI{-2.47}{\femto\meter}\) (chosen arbitrarily) on the unphysical
    sheet adjacent to  the three-body cut.
    The calculation is performed using a rotation angle
    \(\varphi=\SI{0.8}{rad}~ (\ang{45.84})\).
    As a consequence the three-body cut (dashed line) appears at an angle
    \(2\varphi=\SI{1.6}{rad} ~(\ang{91.67})\), close to the negative imaginary
    axis.
    The value where the characteristic polynomial becomes
    zero is indicated by a star, which corresponds to a three-boson resonance.
    Similar plots can be created for the other scattering length values along
    the pole trajectory.}
  \label{fig:3B_Contour}
 \end{figure}

The numerical procedure is as described above. \Cref{fig:3B_Contour} shows a
contour plot of the characteristic polynomial for an arbitrary scattering
length along the contour. The resonance is identified by the zero in the fourth
quadrant.
Based on the Efimov structure of the pole trajectories, we present our results in
units of \(a_-\).
Except for small \(\Lambda\), where we cut off physically relevant momenta, or
for large \(\Lambda\), where our theory is no longer valid,  all Efimov states
are located on top of each other using this unit scheme.

\begin{figure}[htp]
 \includegraphics[width=0.5\textwidth]{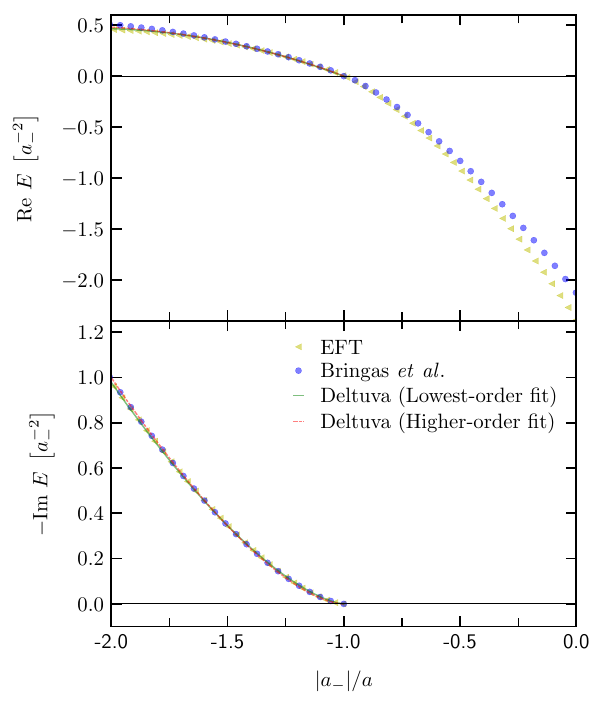}
 \caption{Trajectories for the real (upper) and imaginary (lower) part of the
 three-boson pole energies.
 The trajectories are presented in units of \(a_-\) such that the transition
 from the physical to the unphysical sheet takes places at \(|a_-|/a=-1\).
 The results derived in this work are compared to a calculation using the
 formalism by Bringas \textit{et al.}~\cite{Bringas:2004zz}  and two fits by
 Deltuva~\cite{Deltuva:2020sdd}.
 The latter ones are only present on the unphysical sheet.
 The pole trajectories presented here are equivalent to the trajectory sketched
 in~\cref{fig:ComplexMomPlane} (dashed trajectory) as well
as~\cref{fig:Efimov}.} \label{fig:ThreeBosons}
\end{figure}

This work is compared to the results by Bringas
\textit{et al.}~\cite{Bringas:2004zz} and Deltuva~\cite{Deltuva:2020sdd}.
Similarly to this work, Bringas \textit{et al.} use a Faddeev equation for a renormalized zero-range model and
analytically continue it applying the rotation of the contour of integration.
Deltuva solved the Faddeev equations for the transition operators for several short-range force models on the physical
sheet and matched it to an expansion of the transition operator into a power
series near the resonance pole.
The derived pole trajectory is fitted by a lowest- and a higher-order
approximation.
The results of Bringas \textit{et al.} and Deltuva are shown together with
the pole trajectory derived within this work in \cref{fig:ThreeBosons}.
We only present one pole trajectory as both renormalization prescriptions result
in indistinguishable trajectories.
All results agree very well with our EFT calculation.

A further qualitative comparison is possible with the results by
Jonsell~\cite{Jonsell:2006xx}, who investigates the three-boson system using a
hyperspherical formalism together with the CSM,
and with the work of Hyodo \textit{et al.}~\cite{Hyodo:2013zxa},
who calculate the pole trajectories using a contour rotation. 
The behavior of our pole trajectory is consistent with both these
calculations.

\subsection{Three-neutron system}

After benchmarking our method for the three-boson system,
we focus on the three-neutron system.
The easiest way to analytically continue the
Faddeev equation would be to start with a bound-state pole on the physical
sheet for an unphysical value of the $nn$ scattering length
as for the three-boson system.
However, this method cannot be applied for the three-neutron system
in leading-order pionless EFT because there is no three-body bound state
for any value of the scattering length $a$.

This statement is based on the following argument:
The structure of Eq.~\eqref{eq:3n-Fad} in the limit
$\Lambda\to\infty$ implies that if there is no
three-body bound state for a particular value $a$ of the neutron-neutron
scattering length, then there is no bound state for any other value of
$a$ with the same sign. This argument relies on the fact that,
for $\Lambda\to\infty$, $a$ is the only
dimensionful parameter in the equation and can be scaled out.
The resulting dimensionless equation then applies for any finite value of $a$
with the same sign.
In the case of finite $\Lambda$, one can still exclude all physical
bound states with energies well below the cutoff scale
$|E|\ll \Lambda^2$.
Finite range or other higher-order effects cannot change this
conclusion as long as they are perturbative as stipulated by
the power counting of pionless EFT.
Based on this argument, we have excluded three-neutron bound states
for $\lambda=1$ and $\lambda=0$.\footnote{Higher
values of $\lambda$ were not considered explicitly, but we do not expect
any bound states there, either.}

Since there are no bound states, we use a different ansatz.
Gl\"ockle used the same Faddeev equation for the degenerate channels \(J^\pi=\frac{1}{2}^-\) and \(J^\pi=\frac{3}{2}^-\), together with
a Yamaguchi model \(V_2(p,k) = -\kappa g(p)g(k)\), where
\begin{equation}
 g(p) = \frac{1}{p^2 + \beta^2}
\end{equation}
is a form factor~\cite{Glockle:1978zz}.
Beside the interaction strength \(\kappa\), this form factor implements a
further scale \(\beta\) which induces a finite range.
Using both parameters together it is possible
to reproduce not only the scattering length, but also the effective
range \(r_e\) which is included non-perturbatively.
Keeping \(\beta\) fixed and increasing the interaction strength \(\kappa\),
this allows us to create a three-neutron bound state.
Now, we are able to perform a calculation analogous to the three-boson system.
Starting at a three-neutron bound state and reducing the two-neutron interaction
strength \(\kappa\) while keeping \(\beta\) fixed (increasing the positive
neutron-neutron scattering length) the pole trajectory moves through the
two-body cut onto the unphysical sheet in the third quadrant of the complex
energy plane.
Decreasing \(\kappa\) further, the three- as well as the two-neutron poles
finally arrive at zero.
Following the pionless EFT power counting, we are able to reproduce this
trajectory next to the origin for \(a\) going to infinity.

\begin{figure}[htp] \includegraphics[width=0.45\textwidth]{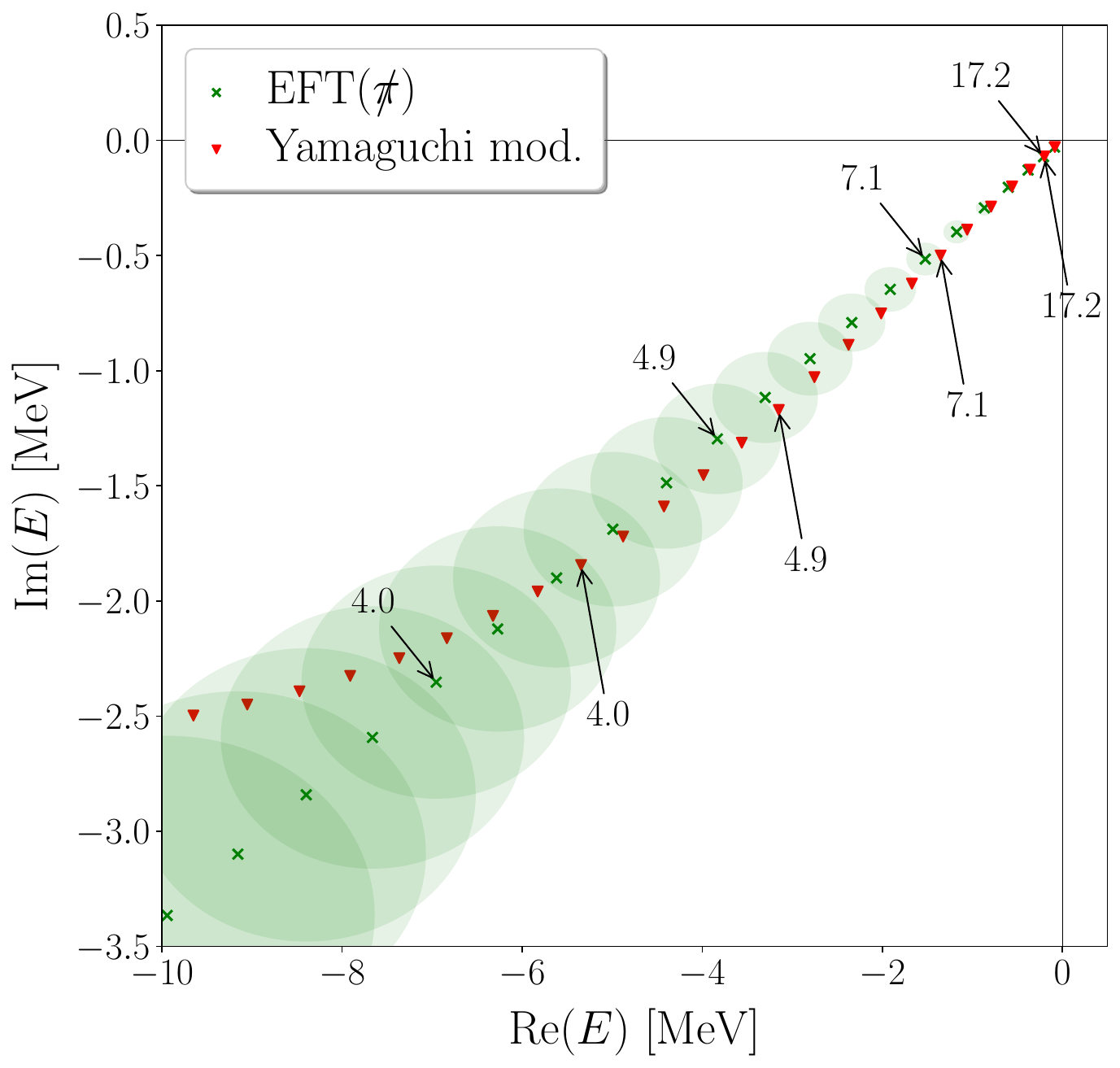}
 \caption{The pole trajectory of the three-neutron system \(( 
  J^\pi=\frac{1}{2}^- ~\mathrm{and}~ J^\pi=\frac{3}{2}^- )\) for positive
  scattering length for a
  pionless EFT interaction together with the LO error bands in comparison to a
  calculation using the Yamaguchi model.
  The values of the scattering length in \si{\femto\meter} for selected points are given by the
  numbers and arrows.
  The pole trajectory calculated using the Yamaguchi model is equivalent to
  the trajectory sketched in \cref{fig:ComplexMomPlane} (solid trajectory).
  While the pionless EFT calculation only includes the scattering length, the
  Yamaguchi model incorporates higher order effective range effects.
  The results are presented in the complex energy plane.
  The lower half-plane shows the unphysical sheet accessed through the
  two-body cut.
  Within the region where EFT\((\slashed{\pi})\) is valid, both trajectories
  agree within the EFT error indicated by the LO circles.
  }
  \label{fig:ThreeNeutrons_Trajectory}
 \end{figure}

\Cref{fig:ThreeNeutrons_Trajectory} shows the part of this trajectory close to
the origin together with LO EFT errors in comparison to the results
derived using the Yamaguchi model.
Along both trajectories the values of the
scattering length (in \si{\femto\meter}) for selected points are indicated.

\begin{figure}[htp]
 \includegraphics[width=0.5\textwidth]{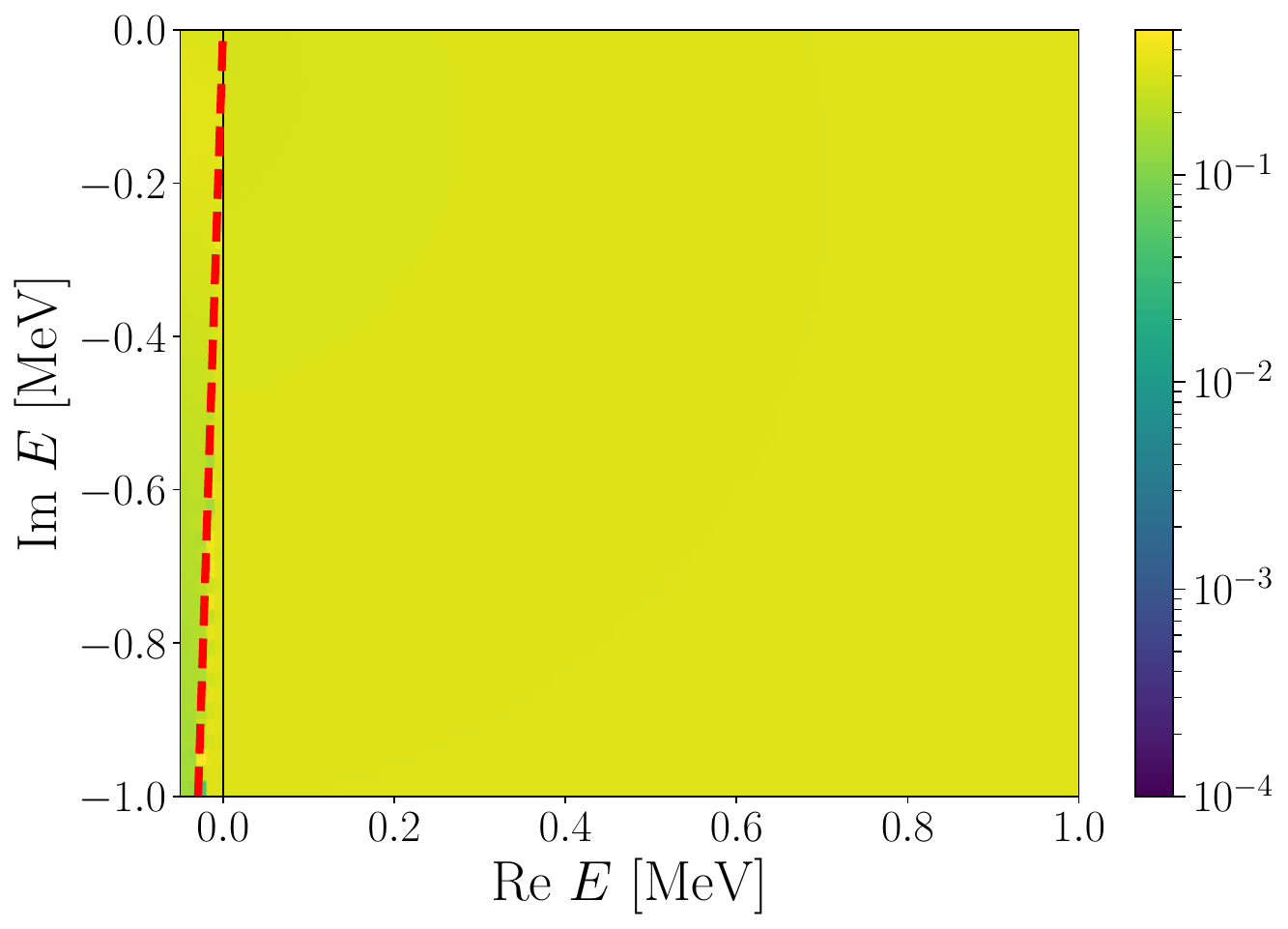}
 \caption{A contour plot presenting the absolute value of the determinant of
 one minus the kernel of Eq.~\eqref{eq:3nContinued} on the unphysical sheet adjacent
 the three-body cut.
 This calculation was performed for \(a=\SI{-18.77}{\femto\meter}\) and a
 rotation angle \(\varphi= \SI{0.8}{rad}\).
 Similarly to \cref{fig:3B_Contour}, the three-body cut is visible at an angle
 \(2\varphi=\SI{1.6}{rad}~(\ang{91.67})\) next to the negative imaginary axis.
 A zero within this plot is equivalent to an eigenvalue equal to 1 of
 Eq.~\eqref{eq:3nContinued}, which itself corresponds to a possible physical state.
 So, this plot presents no evidence for a possible three-neutron resonance in the degenerate \(J^\pi=\frac{1}{2}^-\) and \(J^\pi=\frac{3}{2}^-\) channels.}
 \label{fig:ThreeNeutrons_Contour}
\end{figure}

Now, we arrive at the interesting part of the pole trajectory: an unbound
two-neutron subsystem.
Within Ref.~\cite{Glockle:1978zz}, the pole trajectory continues on an
unphysical sheet which is different from the relevant unphysical sheet next to
the lower rim of the physical sheet.
So, the pole at the physical set of parameters is too far away from the physical
sheet and has no effect on observables. \\
To see if this is different using the Gaussian regulator, we follow the
explanation above and investigate the absolute value of the determinant of
\(\mathbbm{1}\) minus the kernel of Eq.~\eqref{eq:3nContinued} in the complex
momentum plane.
A zero within this plot would indicate a possible pole.
This investigation is performed for negative values of \(a\) starting at
\(\SI{0}{\femto\meter}\) up to the physical value
\(a\approx\SI{-18.9}{\femto\meter}\).
\Cref{fig:ThreeNeutrons_Contour} shows an example of what these contour plots look
like.
Beside the expected discretized cut structures resulting from the structure of
the equation, no behavior that can be connected to a pole is visible.
So, using a Gaussian regulator we recover the results of
Ref.~\cite{Glockle:1978zz}.
This outcome is also supported by the power counting, predicting that we should
be able to recover the results of the Yamaguchi type regulator for large
negative scattering lengths.

Similarly to the negative parity channels discussed above, we have investigated the analytically continued Faddeev equation for the \(J^\pi=\frac{1}{2}^+\) channel (\(\lambda=0\)) in the complex momentum plane in the vicinity of the physical \(nn\) scattering length. We found no evidence of a three-neutron resonance in the \(J^\pi=\frac{1}{2}^+\) channel, either.

\section{Finite volume}
\label{sec:FV}

Finite-volume calculations were established in Ref.~\cite{Klos:2018sen} as
a tool to identify few-body resonance states.
This approach goes back to the pioneering work of
Lüscher~\cite{Luscher:1986pf,Luscher:1990ux}, who first showed that properties
of the infinite-volume \(S\) matrix, and therefore observables like bound states
and scattering parameters, are encoded in how the discrete energy levels in a
finite periodic box change as the size $L$ of the box is varied.
Resonance states are manifest as avoided crossings of energy levels, which is
well established for two-body
systems~\cite{Wiese:1988qy,Luscher:1991cf,Rummukainen:1995vs} and 
routinely used to extract resonance properties from 
lattice QCD calculations \cite{Briceno:2017max}.
In Ref.~\cite{Klos:2018sen}, it was demonstrated that this signature carries over to the few-body sector.
We use here these findings, and in particular the discrete variable
representation (DVR), as an additional, independent tool to search for
three-neutron resonances.

\subsection{Basic setup}

The starting point for the DVR construction of states in a periodic box with
edge length $L$ are plane-wave states $\phi_j(x)$
with $j={-}n/2,\cdots n/2-1$ for even $n>2$, where $x$ denotes the relative
coordinate describing a two-body ($N=2$) system in one dimension ($d=1$).
Any periodic solution of the one-dimensional (1D) Schrödinger equation can be expanded in terms of
the states $\phi_j(x)$, yielding a discrete Fourier transform (DFT).
Given a set of equidistant points $x_k \in [{-}L/2,L/2)$ and weights $w_k = L/n$
(independent of $k$), DVR states are constructed as~\cite{Groenenboom:2001web}
\begin{equation}
 \psi_k(x) = \dvrsum{i}{n} \mathcal{U}^*_{ki} \phi_i(x) \,,
\label{eq:psi-dvr}
\end{equation}
with $\mathcal{U}_{ki} = \sqrt{w_k} \phi_i(x_k)$ defining a unitary matrix.
The DVR is convenient for two main reasons:
\begin{enumerate}
 \item For a local interaction, the potential operator ${V}$ reduces
  (approximately) to a diagonal matrix, $\mbraket{\psi_k}{{V}}{\psi_l}
  \approx V(x_k) \delta_{kl}$, where the quality of this approximation is
  determined by the number $n$ of discretization points.
  This holds for any number $d$ of spatial dimensions and $N$ interacting
  particles.
 \item The kinetic energy ${K}$ is \emph{not} diagonal, but its form is
  known analytically:
\begin{equation}
 \mbraket{\psi_k}{{K}}{\psi_l}
 = \begin{cases}
  \dfrac{\pi^2 N^2}{6\mu L^2} \left(1+\dfrac{2}{n^2}\right)\,
  &\text{for}\ k = l\,, \\[0.5em]
  \dfrac{({-}1)^{k-l} \pi^2}{\mu L^2\sin^2\!\big(\pi(k-l)/n\big)}\,
  &\text{otherwise}\,.
 \end{cases}
\label{eq:T-DVR-1D}
\end{equation}
  For $d>1$ or $N>2$ the DVR representation of ${K}$ becomes a sparse matrix
  that can be calculated very efficiently based only on the $d=1$, $N=2$
  elements.
  Alternatively, as pointed out in Ref.~\cite{Bulgac:2013mz}, one can exploit
  the relation of the plane-wave based DVR to the DFT and evaluate the kinetic
  energy in momentum space, but we find the direct calculation more efficient
  for large-scale calculations.
\end{enumerate}

The construction is straightforward to generalize to the case of an arbitrary
number of particles $N$ and spatial dimensions $d$, starting from product states
of $(N-1)\times d$ plane waves, one for each relative-coordinate component.
The transformation matrices and DVR basis functions are defined \via tensor
products, and DVR states are labeled by a collection of $(N-1)\times d$ indices.
Using the shorthand notation $\ket{\psi_k} = \ket{k}$, a general state is
written as
\begin{multline}
 \ket{s}
 = \ket{
  (k_{{1,1}},\cdd, k_{{1,d}}),\cdd ,(k_{{N-1,1}},\cdd, k_{{N-1,d}});
  (\sigma_1,\cdd, \sigma_N)
 } \,,
\label{eq:s}
\end{multline}
where the $\sigma_i$ denote the spin degrees of freedom.
The space spanned by all these states $\ket{s}$ is denoted by $B$.
For spinless bosons, $\sigma_i=0$ for all $i$, whereas in general, for particles
with spin $S$, $\sigma_i={-S},\cdot\cdot S$.
Specifically, we have here $S=1/2$, $d=3$, and $N=3$ for the three-neutron system
we study.

Fore more details regarding the DVR setup we refer to
Refs.~\cite{Klos:2018sen,Konig:2020lzo} and further references cited therein,
and merely recall here that the overall strategy with this method is to
represent the $N$-body finite-volume Hamiltonian $H = H_0 + V$, where $H_0 =
\hat{K}$ is the relative kinetic energy operator and $V$ denotes the sum of all
interactions among the particles, as a matrix in the space spanned by the DVR
states.
Energy levels in the box are then obtained by calculating the spectrum (or
more specifically the lowest lying states in the spectrum) via Lanczos/Arnoldi
iteration.

\subsection{Separable interactions}

The DVR as described above is set up only to work with local potentials.
In order to study in finite volume the same separable potentials as used for the
momentum-space calculation, we discuss in the following the appropriate
extension of the DVR formalism.
We start from the definition of a general DVR state,
Eq.~\eqref{eq:s}, and neglect for the moment the spin degrees of freedom.

Recall that projected onto coordinate space a state $\ket{s}$ is a product of
one-dimensional DVR wave functions,
\begin{equation}
 \psi_s(\underline{x})
 = \braket{\underline{x}}{s}
 = \prod_{
  \substack{i=1,\cdot\cdot N-1 \\c=1,\cdot\cdot d}
 } \psi_{k_{i,c}}(x_{i,c}) \,,
\end{equation}
where $\underline{x}$ is used to denote the collection of all relative
coordinates.
Furthermore, let $\psi(\underline{x})$ be a generic state expanded in the DVR
basis $B$,
\begin{equation}
 \psi(\underline{x}) = \sum_{s\in B} c_s \psi_s(\underline{x}) \,.
\label{eq:psi_s}
\end{equation}
This could be an actual eigenstate of the Hamiltonian, or any intermediate state
vector that is encountered during the Lanczos-based diagonalization of the
Hamiltonian.
Either way, the $\{c_s\}$ is a finite vector of coefficients with entries as
introduced in Eq.~\eqref{eq:s}.

Let us now consider a (rank-1) separable two-body potential, generically written
in coordinate space as
\begin{equation}
 V_2(x,x') = C \, g(x) g(x') \,,
\label{eq:V-sep}
\end{equation}
where $C$ is the strength and $\ket{g}$ with $\braket{x}{g} = g(x)$ is the
``form factor.''
For simplicity we restrict the following discussion to one spatial dimension
since everything carries over to $d>1$ in a straightforward way.
For a two-body state, we have $\underline{x} = x_1 \equiv x$ and $\ket{s} =
\ket{k}$ (a single index describes the spinless 1D state), so that applying $V$
is straightforward:
\begin{spliteq}
 \mbraket{s}{V_2}{\psi}
 &= \int\!\dd x \int\!\dd x' \, \psi_{s}^*(x') V_2(x,x')
 \sum_{s'\in B} c_{s'} \psi_{s'}(x') \\
 &= C \, g(x_{s}) \sum_{s'\in B} c_{s'} \, g(x_{s'}) \,.
\label{eq:s-V-psi}
\end{spliteq}
We have used here the DVR property of the states, and $x_s = x_k$ is the
location of the lattice point characterizing the state $\ket{s}$.
It follows, that for this case, applying a separable potential to a generic DVR
state is very simple, and the factorization property of the potential is
directly reflected in the end result.

For more than two particles one (in general) needs to consider the separable
potential~\eqref{eq:V-sep} between all pairs of particles.
Since for each pairwise interaction the potential only involves the relative
coordinate of that pair, appropriate delta functions need to be included for all
``spectator'' particles that do not take part in the particular interaction.
The way to do this consistently across all pairs is to express these delta
functions such that they fix the position of the spectators relative to the
center of mass of the interacting pair (which amounts to a partial
transformation to a particular set of Jacobi coordinates, similar to what is
used in the Faddeev formalism discussed in the main text).

A minor complication arises from the fact that in a periodic finite volume the
center of mass of a cluster of particles is not uniquely defined.
This can already be seen for two particles in one dimension:
consider these particles on a periodic interval from 0 to $L$ residing at
positions $x=1$ and $x=L-1$; then both $x=0$ and $x=L/2$ are valid candidates
for the particle's center of mass.
Visualizing the periodic interval as a circle, these two possibilities
correspond to the middle points of the two arcs that connect the particles.
With increasing number of dimensions and number of particles, the set of
candidates for the center of mass becomes larger.
Which one is chosen is arbitrary, but the choice has to be consistent.
To that end, for a configuration of $A$ coordinates $C=\{\vecx_i\}_{i=1}^A$
we define the center of mass to be that point that minimizes the sum
of distances of all particles measured with respect to the center of mass:
\begin{equation}
 \vecR_\cm = \argmin_{\vecR\in S_\cm(C)}\left(
  \sum_{j=1}^A d_L(\vecR,\vecx_j)^2
 \right) \,,
 \label{eq:R-cm}
\end{equation}
where $S_\cm(C)$ is the set of all possible center-of-mass coordinates for the
given configuration $C$ and $d_L$ measures the distance between two points as
the shortest path between them while accounting for the periodic boundary
condition.

Taking into account the spatial lattice nature of the plane-wave DVR basis and
noting that for an $A$-body state on a lattice of extent $n$ the center of mass
falls onto a lattice with extent $nA$~\cite{Elhatisari:2017eno}, it becomes
straightforward to express the center-of-mass coordinate as an index in an
enlarged DVR space.
For each given pair interaction $V_{ij}$, denoting a potential of the
form~\eqref{eq:V-sep} acting between particles $i$ and $j$, such an index is
considered for each spectator particle in order to include appropriate
Kronecker deltas.
Schematically, Eq.~\eqref{eq:s-V-psi} becomes:
\begin{equation}
 \mbraket{s}{V_{ij}}{\psi}
 = \mathcal{N} \times C \, g(x_{s;ij}) \;
  \sum_{
   \mathclap{\substack{s'\in B \\ r_{s';k,ij} = r_{s;k,ij} \forall k \neq i,j}}
   } \;
   c_{s'} \, g(x_{s';ij})
   \rule{0pt}{2em}
   \,.
\label{eq:s-V-psi-ij}
\end{equation}
Here $x_{s;ij}$ denotes the relative distance (modulo the periodic boundary)
between particles $i$ and $j$ in configuration $\ket{s}$ and $r_{s;k,ij}$
is the coordinate of particle $k$ relative to the center of mass of particles
$i$ and $j$ as defined in Eq.~\eqref{eq:R-cm}, for $A=2$.
The generalization to $d>1$ is trivial, and only a minor technical complication
arises from the fact that the DVR states are expressed in relative coordinates.
For $j=N$ one can directly work with the $x_i$ (or $x_{i,c}$ in $d>1$),
whereas for other pairs one considers appropriate differences of the $x_i$ that
give the desired coordinate vector.
The factor $\mathcal{N} = 2^d({L}/{N})^{d/2}$ arises as normalization from the
integral over spectator coordinates.

Since the plane-wave DVR states we consider are closely related to a
DFT~\cite{Bulgac:2013mz,Konig:2020lzo}, this is naturally the tool to
use for calculating matrix elements $\mbraket{s}{V_{ij}}{\psi}$ when the
$V_{ij}$ are given in momentum space.
To that end, one considers two-body momentum modes
\begin{equation}
 p_j = \frac{2\pi j}{L}
\end{equation}
and, for the appropriate coordinate $x_{s;ij}$ as introduced above, evaluates
\begin{equation}
 \braket{s}{g}
 = \sum_{k={-}n/2}^{n/2} g(p_k) \, \exp(\ii p_k x_{s;ij}) \,.
\label{eq:g-sep-ms}
\end{equation}
This equation is written for the one-dimensional case, but it straightforwardly
generalizes to $d>1$.
The considerations about including appropriate delta functions to fix the
coordinates of the spectator particles relative to the interacting pair's center
of mass remain exactly the same and need not be carried out in momentum space.

The momentum-space implementation also makes it particularly convenient to
consider interactions that act only in a single partial wave.
The following considerations can easily be generalized to arbitrary $d$, but we
consider here only the most relevant case $d=3$.
For spinless particles, the projection is achieved by merely including a factor
$|\vec{p}_{\vec{k}}|^l Y_{lm}(\hat{\vec{p}}_{\vec{k}})$ in the three-dimensional
generalization of Eq.~\eqref{eq:g-sep-ms}, where $\vec{p}_{\vec{k}}$ is a
momentum mode in 3D and $Y_{lm}$ denotes the
spherical harmonic for angular momentum $l$ and projection $m$.\footnote{%
Note that $|\vec{p}_{\vec{k}}|^l Y_{lm}(\hat{\vec{p}}_{\vec{k}})$ really is a
solid harmonic in momentum space.} Note that since the DVR uses a full
three-dimensional model space (not decomposed into partial waves), a potential
term needs to be included for each $m={-}l,\cdot\cdot, l$.
To include spin, one can directly utilize the projection indices $\sigma_k$
included in the states~\eqref{eq:s}.
If the interaction is meant to act in a two-body channel ${}^{2s+1}l_j$, written
in spectroscopic notation, one includes a Clebsch-Gordan coefficient that
couples the individual particles spins to $s$ (where the total projection is
fixed to be $\sigma_i+\sigma_j$), and then another Clebsch-Gordan coefficient
that couples $l$ and $s$ to total angular momentum $j$.
For this case, one has a potential for each $m_j={-}j,\cdot\cdot, j$, and all
projection quantum numbers in the Clebsch-Gordan coefficients are fully
determined by this, the spin projection, and the standard Clebsch-Gordan
selection rules.
This means that there are no extra sums required to carry out the partial-wave
projection.
In essence, this sum is the one appearing already in Eq.~\eqref{eq:s-V-psi-ij}.

For practical implementations it is desirable to avoid complex arithmetic as
much as possible.
To achieve that, it is convenient to work with real spherical
harmonics instead of the $Y_{lm}$ and replace in Eq.~\eqref{eq:g-sep-ms} the
exponential function with a cosine or sine for even and odd $l$, respectively.

\subsection{Results}

For EFT applications it is convenient to express the separable
potential~\eqref{eq:V-sep} in momentum space as
\begin{equation}
 V_2(q,q') = C \, g(q) g(q') \,,
 \label{eq:V-sep-q}
\end{equation}
where $g(q)$ is the Fourier transform of $g(r)$, indicated only by the argument
for simplicity.
In typical applications the potential is often given directly in
the form~\eqref{eq:V-sep-q}.
We pick here specifically a
super-Gaussian form:
\begin{equation}
 g(q) = \exp({-}q^{2n} / \Lambda^{2n}) \,.
\end{equation}
The Faddeev calculations discussed in Sec.~\ref{sec:eft} use this form with
$n=1$.
To ensure that our finite-volume implementation of separable interactions is
correct, we have run bound-state benchmark calculations with such simple
Gaussian form factors for some selected potentials.
For the three-neutron results discussed in the following, however, we chose to
work with $n=2$ because  this regulator form provides a stronger
suppression of high-momentum modes, which helps to achieve converged
calculations in large boxes.
We moreover chose a rather soft cutoff scale $\Lambda = 250~\MeV$ for these
calculations.

\begin{figure}[htp]
 \includegraphics[width=0.5\textwidth]{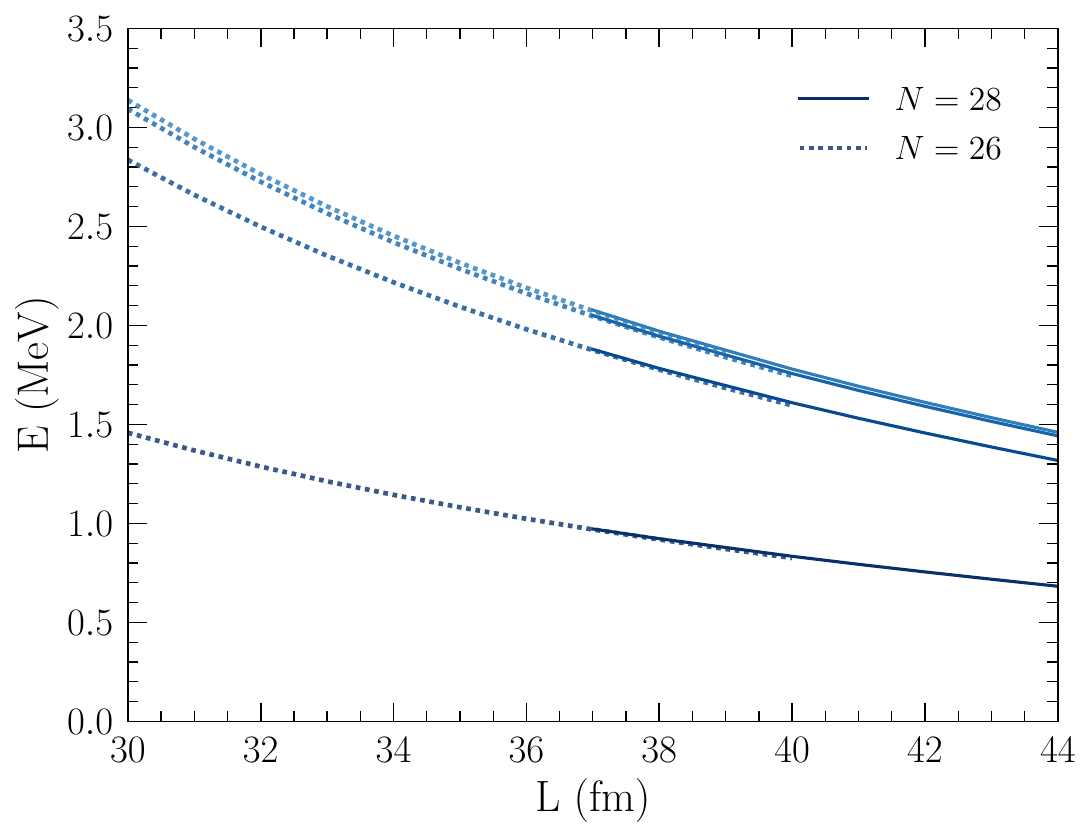}
 \caption{Finite-volume energy levels for three neutrons with total spin $S=1/2$
  and negative parity, calculated assuming a separable $n=2$ super-Gaussian
  interaction tuned to reproduce a neutron-neutron scattering length $a =
  {-}18.9~\fm$.
  The solid (dashed) lines were obtained using $N=26$ ($28$) mesh points to
  construct the three-neutron DVR basis.
 \label{fig:FVphys}
 }
\end{figure}

\begin{figure}[htp]
 \includegraphics[width=0.5\textwidth]{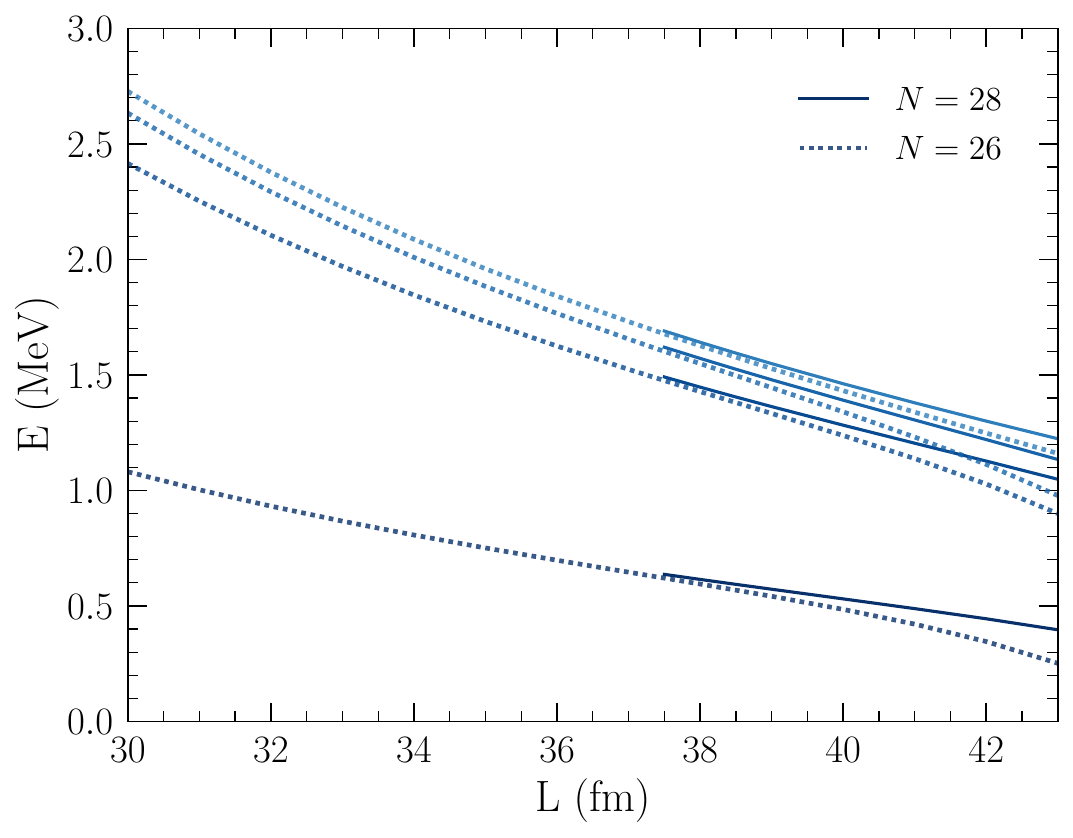}
 \caption{Same as Fig.~\ref{fig:FVphys}, except with the interaction tuned
  to a neutron-neutron scattering length $a = {+}18.9~\fm$.
 \label{fig:FVplus}
 }
\end{figure}

\begin{figure}[htp]
 \includegraphics[width=0.5\textwidth]{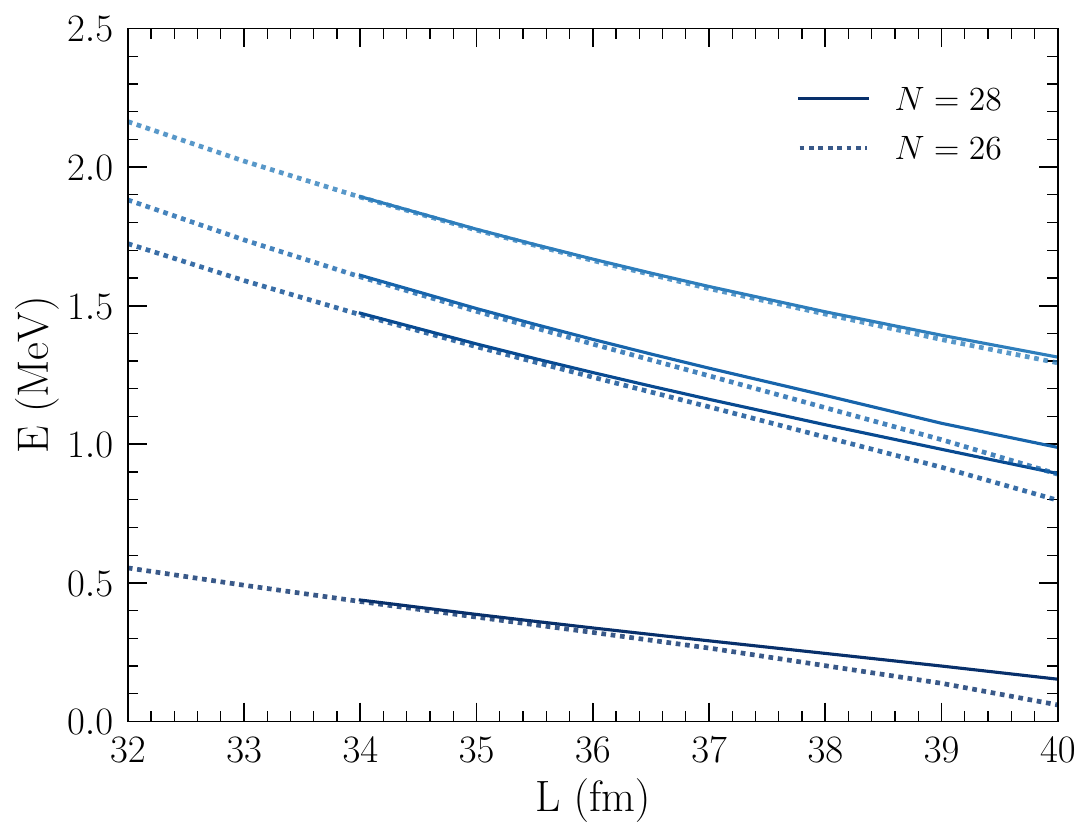}
 \caption{Same as Fig.~\ref{fig:FVphys}, except with the interaction tuned
  to a neutron-neutron scattering length $a = {+}10.0~\fm$.
 \label{fig:FVplus10}
 }
\end{figure}

Results are shown in Figs.~\ref{fig:FVphys}, \ref{fig:FVplus},
and~\ref{fig:FVplus10} for 
total spin $S=1/2$ and negative parity.
A projection on the representations of the cubic symmetry
group of the box shows that these energy levels correspond to
the degenerate \(J^\pi=\frac{1}{2}^-\) and \(J^\pi=\frac{3}{2}^-\) channels we are primarily interested in. 
We start with the physical neutron-neutron scattering
length (Fig.~\ref{fig:FVphys}) and then gradually adjust the interaction to
become more attractive.
In particular, for $a = {+}18.9~\fm$ (Fig.~\ref{fig:FVplus}) and
$a = {+}10.0~\fm$ (Fig.~\ref{fig:FVplus10}) the interaction supports shallow
dineutron bound states with energy $E_2 = 1/(M_n a^2)$.
In the three-body spectrum we can see the effect of the increasing attraction
directly reflected in the fact that all energy levels get shifted downwards
in going from Fig.~\ref{fig:FVphys} to Fig.~\ref{fig:FVplus10}.
The comparison of different DVR basis size (solid and dashed lines in the
figures) shows that we can achieve sufficiently converged calculations in the
energy range where three-neutron resonances have been speculated to exist.
However, for the values of the scattering length considered, we do not see
avoided level crossings in the spectrum or plateau shapes in individual
energy levels. Thus, we see none of the signatures of a resonance discussed in Ref.~\cite{Klos:2018sen}.
Our finite-volume results thus confirm the findings of our Faddeev calculations that there are no resonances in the degenerate \(J^\pi=\frac{1}{2}^-\) and \(J^\pi=\frac{3}{2}^-\) channels,
and we conclude that even with increased attraction in the neutron-neutron
\(S\)-wave interaction there is no support for a three-neutron resonance state
in the most likely channel.
For the physical neutron-neutron scattering length we have furthermore calculated a positive-parity spectrum (not shown as an explicit figure), in which we see no indications for a three-neutron resonance either.

\section{Conclusion and outlook}
\label{sec:conc}

In this work, we have presented a rigorous study of the appearance of resonances
in three-body systems using two complementary methods.
We first use pionless effective field theory at leading order to write down an
effective interaction potential which may include a three-body interaction.
In the second step, we analytically continue the Faddeev equation in momentum
space to the unphysical sheet adjacent to the positive real energy axis using a
rotation of the integration contour.
On the unphysical sheet, we search for poles corresponding to resonances and
virtual states.
This direct search for resonance poles is complemented by  an alternative
finite-volume method based on identifying avoided crossings of energy levels as
the size of the volume is varied.
The two methods have very different systematics and ideally complement each
other, although the latter method is not suitable for virtual states.

We apply our framework to two types of systems: (i) three bosons with large
negative scattering length $a$ and (ii) the three-neutron system.
Our study of the three-boson system also serves as a test case of our method.
It is well known that three-body Efimov states for $a<0$ turn into resonances as
they cross the three-particle threshold.
We confirm the previous calculations by Bringas~\etal~\cite{Bringas:2004zz}
and Deltuva~\cite{Deltuva:2020sdd} both qualitatively and quantitatively.
Moreover, our results are qualitatively consistent with the pole trajectories of
Jonsell~\cite{Jonsell:2006xx} and Hyodo \textit{et al.}~\cite{Hyodo:2013zxa}.
The trajectories of the Efimov resonances can be used to explain the behavior of
the three-body recombination rate of three spinless bosons at low temperatures
\cite{Jonsell:2006xx,Yamashita_2007}.

The main motivation for our work is the suggestion of a low-energy resonance or
virtual state in the three-neutron system~\cite{Gandolfi:2016bth,Li:2019pmg}.
We can reproduce earlier calculations by Gl\"ockle for a three-neutron Yamaguchi
model system with a strong attraction~\cite{Glockle:1978zz}.
However, we do not find any resonances for the physical case in the relevant
$\lambda=1$ and $\lambda=0$ orbital angular momentum channels 
corresponding to the $J^\pi=\frac{1}{2}^-$, $\frac{3}{2}^-$, and $\frac{1}{2}^+$ channels
in the analytical continuation framework. 
Low-energy resonances in the negative parity channels
are also excluded in the finite-volume framework.
Using the analytical continuation method, we also exclude a three-neutron virtual state.
Our model-independent result agrees with several other recent theoretical
studies~\cite{Deltuva:2018lug,Ishikawa:2020bcs,Higgins:2020avy,Higgins:2020pbe} and
rules out the possibility of a three-neutron resonance or virtual state at low
energy.
Although we use pionless EFT at leading order, we
expect our result to hold also in the presence of higher-order interactions. In pionless EFT
the higher-order terms, including the effective range $r_0$ and $P$-wave
interactions, are purely perturbative and cannot produce any new poles.
Thus the existence of a low-energy three-neutron resonance would also imply the
breakdown of pionless EFT in the three-neutron system.

Obviously, our study does not address the question of four-neutron resonances.
Experimental evidence for such a four-neutron resonance was recently presented
in \cite{Kisamori:2016jie}; see also Ref.~\cite{Marques:2021mqf}
for a current review of the field. We leave this question for
future work.

Finally, we stress that multi-neutron energy spectra contain much
interesting physics, even if multi-neutron resonances are not observed in
experiment.
In Ref.~\cite{Hammer:2021zxb}, e.g., it was pointed out that the
multi-neutron spectra for center-of-mass energies $E$ in the range
$1/(m a^2) \approx \SI{0.1}{\mega\electronvolt} \ll E \ll  1/(m r_0^2)
\approx \SI{5}{\mega\electronvolt}$ are determined by conformal symmetry
up an overall normalization.
Conformal symmetry implies that the multi-neutron correlation functions have
only cuts but no poles, which is consistent with our results for the
three-neutron system.
The neutron spectra show power-law behavior with, in general, fractional
exponents determined by the scaling dimension of the corresponding conformal
field operators.
This is markedly different from weakly interacting particles. Neutron resonance
experiments are ideally suited to confirm this prediction.

\begin{acknowledgments}
We thank Joel Lynn for collaboration in the early stages of this work, and Dean
Lee for useful discussions regarding the implementation of separable
interactions in finite volume.
This work was supported in part by the Deutsche Forschungsgemeinschaft (DFG,
German Research Foundation) -- Project-ID 279384907, SFB 1245, by the BMBF
Contracts No.~05P18RDFN1 and 05P21RDFNB, and by the National Science Foundation
under Grant No.~PHY--2044632.
This material is based upon work supported by the U.S.\ Department of Energy,
Office of Science, Office of Nuclear Physics, under the FRIB Theory Alliance
award DE-SC0013617.
Computational resources for parts of this work were provided by the Jülich
Supercomputing Center.
\end{acknowledgments}

\end{document}